\documentclass[lettersize,journal]{IEEEtran}
\usepackage{amsmath,amsfonts,amsthm}
\usepackage{algorithmic}
\usepackage{algorithm}
\usepackage{array}
\usepackage[caption=false,font=normalsize,labelfont=sf,textfont=sf]{subfig}
\usepackage{textcomp}
\usepackage{stfloats}
\usepackage{url}
\usepackage{verbatim}
\usepackage{graphicx}
\usepackage{cite}
\usepackage[utf8]{inputenc} 
\usepackage[T1]{fontenc}    
\usepackage{hyperref}       
\usepackage{url}            
\usepackage{booktabs}       
\usepackage{amsfonts}       
\usepackage{nicefrac}       
\usepackage{microtype}      
\usepackage{xcolor}         
\usepackage{graphicx}
\usepackage{subfig}
\usepackage{amssymb}
\usepackage{color}
\usepackage{xcolor}
\usepackage{caption}

\def\0{{\mathbf 0}}
\def\1{{\mathbf 1}}

\def\c{{\mathbf c}}

\def\e{{\mathbf e}}
\def\f{{\mathbf f}}

\def\r{{\mathbf r}}

\def\u{{\mathbf u}}
\def\v{{\mathbf v}}
\def\w{{\mathbf w}}
\def\x{{\mathbf x}}
\def\y{{\mathbf y}}
\def\z{{\mathbf z}}

\def\A{{\mathbf A}}

\def\D{{\mathbf D}}

\def\I{{\mathbf I}}

\def\L{{\mathbf L}}
\def\M{{\mathbf M}}

\def\P{{\mathbf P}}

\def\R{{\mathbf R}}
\def\S{{\mathbf S}}
\def\T{{\mathbf T}}

\def\V{{\mathbf V}}
\def\W{{\mathbf W}}

\def\ie{{\textit{i.e.}}}
\def\eg{{\textit{e.g.}}}

\def\cA{{\mathcal A}}

\def\cC{{\mathcal C}}

\def\cE{{\mathcal E}}

\def\cG{{\mathcal G}}

\def\cK{{\mathcal K}}
\def\cL{{\mathcal L}}

\def\cO{{\mathcal O}}

\def\cT{{\mathcal T}}

\def\cV{{\mathcal V}}

\def\bnu{{\pmb{\nu}}}

\newcommand{\green}[1] {\textcolor[rgb]{0.0,0.75,0.0}{{#1}}}

\newtheorem{theorem}{Theorem}
\newtheorem{lemma}{Lemma}
\newtheorem{corollary}{Corollary}

\begin{document}


\title{Deep Unrolling of Lanczos-based Laplacian Eigensolver for \\
Crop Classification in Satellite Images}


\author{%
	 Parham Eftekhar, %
	Gene Cheung,~\IEEEmembership{Fellow, IEEE} %
    \thanks{Gene Cheung acknowledges the support of the NSERC grant RGPIN-2025-06252.  (Corresponding author: Parham Eftekhar.)}
    \thanks{Parham Eftekhar and Gene Cheung are with the Department of EECS, York University, Toronto, M3J 1P3, Ontario, Canada (e-mail: \{eftekhar,genec\}@yorku.ca).}
}

\markboth{}%
{Shell \MakeLowercase{\textit{et al.}}: A Sample Article Using IEEEtran.cls for IEEE Journals}


\maketitle

\begin{abstract}
Per-pixel crop classification in satellite images is an important pre-processing step in remote crop monitoring. 
Leveraging progress in signed graph signal processing, we build an interpretable and lightweight neural network for binary crop classification (e.g., corn versus non-corn) in two stages: i) learn a signed graph from data, and ii) unroll a Lanczos-based eigensolver to compute the eigenvector $\u_1$ associated with the smallest eigenvalue of a hybrid graph Laplacian $\cL$ for spectral clustering. 
Specifically, we first construct a near-balanced signed graph $\cG^s$, where signed edge weights $w_{i,j}\in(-1,1)$ are computed from learned feature distances $d_{i,j}$ via a hyperbolic tangent mapping. 
Given $\cG^s$, we estimate $\u_1$ by unrolling the Lanczos algorithm and use $\mathrm{sign}(\u_1)$ for binary classification. 
Mathematically, we develop a theorem establishing conditions under which $\u_1$ can be well approximated after finite  Lanczos iterations. 
Motivated by this theorem, we relax the classical Lanczos orthogonality constraints during unrolling and optimize end-to-end classification performance while promoting the derived conditions for accurate eigenvector approximation. 
Experiments on real-world Harmonized Landsat-Sentinel imagery show classification performance comparable to state-of-the-art deep learning models while reducing parameters by up to two orders of magnitude.
\end{abstract}

\begin{IEEEkeywords}
Signed graph signal processing, algorithm unrolling, Lanczos algorithm, spectral clustering.
\end{IEEEkeywords}

\section{Introduction}
\label{sec:intro}

\textit{Crop classification} \cite{craig2010history} in satellite images is the problem of classifying each pixel into one of two or more crop types growing on an observed agricultural land; it is an important pre-processing step in remote crop monitoring \cite{monitoring}. 
In the United States, agricultural production is dominated by a small number of major crops, including corn and soybean. 
Consequently, binary classification tasks, such as corn versus non-corn, are particularly relevant in major crop-producing regions, such as Iowa in the US Corn Belt \cite{green2018usa}.
In this paper, we focus on the \textit{binary crop classification} problem, where each major crop is distinguished from all other crop types (\eg, corn vs. non-corn). 
Multi-crop classification can, in turn, be constructed systematically using binary classifiers as building blocks \cite{allwein2000reducing}.

\begin{figure*}[t]
    \centering
    \includegraphics[width=0.7\textwidth]{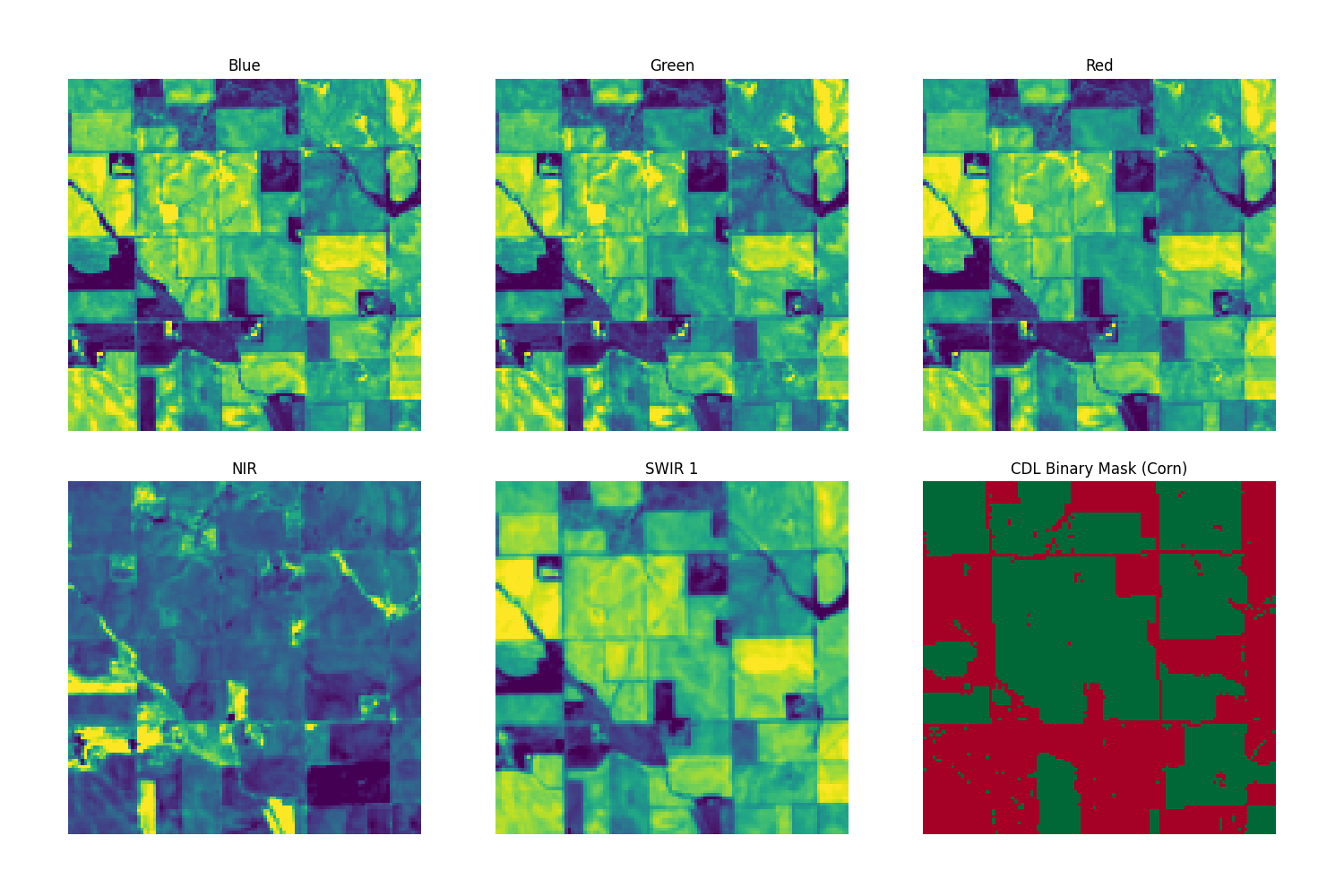}
    \vspace{-0.25in}
    \caption{Example satellite image showing the Blue, Green, Red, NIR, and SWIR-1 bands alongside the corresponding binary mask for corn.}
    \label{fig:all_channels}
\end{figure*}

Unlike natural images, which typically comprise three color channels (\ie, Red, Green, Blue), satellite images capture per-pixel responses across multiple spectral bands \cite{richards2022remote}. 
Different crop types exhibit distinct spectral signatures across these bands, providing discriminative information for crop identification. 
The goal of crop classification is therefore to learn relevant spectral representations from labeled image data in a supervised manner, so that the learned representations can be used to distinguish crop types on a per-pixel basis in unseen satellite images. 
Consequently, crop classification can be naturally formulated as a semantic segmentation task \cite{rustowicz2019semantic}. 
Fig.\;\ref{fig:all_channels} shows an example of the spectral-band imagery alongside its corresponding corn mask derived from the Cropland Data Layer (CDL) \cite{boryan2011monitoring}.

Recent advances in deep learning have led to highly effective semantic segmentation architectures \cite{long2015fully}, including FCNs, DeepLab, PSPNet, and their successors \cite{li2018pyramid,UPerNet,MAnet}. 
By incorporating increasingly sophisticated mechanisms such as multi-scale feature aggregation and attention \cite{zhao2017pyramid,chen2019rethinking}, these models achieve impressive segmentation performance. 
However, they are typically very large in size, requiring millions of parameters, and are difficult to interpret.

Motivated by these limitations, we leverage recent advances in signed graph frequency analysis \cite{yang22,dinesh24} within the \textit{graph signal processing} (GSP) field \cite{ortega18ieee,cheung18} to develop an end-to-end signed graph spectral network for binary crop classification. 
Our network employs a two-stage process. 
In the first stage, the \textit{signed graph learning} stage, we compute a learned feature representation $\f_i \in \mathbb{R}^K$ for each pixel $i$ of $N$ total pixels through a lightweight \textit{convolutional neural network} (CNN) \cite{lecun1998gradient,goodfellow2016deep}, using which we map feature distances $d_{i,j} \in \mathbb{R}_+$ between pixel pairs $(i,j)$ to \textit{signed} edge weights $w_{i,j} \in (-1,1)$ via a hyperbolic tangent mapping, resulting in a signed graph $\cG^s$. 
Unlike conventional positive graphs $\cG^+$ in GSP \cite{ortega18ieee,cheung18} that encode only pairwise similarities of varying degrees, we contend that a large feature distance $d_{i,j}$ implies large dissimilarity between the associated pixel-pair $(i,j)$, and thus a negative edge weight $w_{i,j} \approx -1$ promoting opposite label assignment is essential. 
Appropriate feature learning thus roughly clusters the $N$ pixels into two groups in feature space, where two pixels $i$ and $j$ within the same group have small distance $d_{i,j} \approx 0$ (and thus positive edge weight $w_{i,j} > 0$), while two pixels $i$ and $l$ from different groups have large $d_{i,l} \gg 0$ (and thus negative edge weight $w_{i,l} < 0$), resulting in a \textit{near-balanced}\footnote{By the Cartwright-Harary's Theorem \cite{harary1953notion}, a signed graph $\cG^s$ is balanced iff a  positive (negative) edge always connects two nodes of the same (different) group. See Section\;\ref{subsec:graph_balance} for details.  } graph $\cG^s$.
For a balanced signed graph $\cG^s$, it was shown \cite{yang22} that the sign pattern of the smallest-eigenvalue eigenvector exactly recovers the underlying two-way partition, motivating the use of $\mathrm{sign}(\u_1)$ for binary classification.

In the second stage, the \textit{eigen-solver} stage, we compute the first eigenvector $\u_1 \in \mathbb{R}^N$ corresponding to the smallest eigenvalue $\lambda_{\min}(\cL)$ of a hybrid graph Laplacian $\cL \in \mathbb{R}^{N \times N}$ by building a fixed iteration number $m$ of the Lanczos algorithm \cite{lanczos1950iteration} as neural layers in a sequence, leveraging the \textit{algorithm unrolling} paradigm \cite{monga2021algorithm}. 
The key insight is that,
since the learned signed graph $\cG^s$ is typically only near-balanced rather than perfectly balanced, $\mathrm{sign}(\u_1)$ may not exactly recover the optimal class partition.
Thus, we unroll the Lanczos iterations and relax the classical orthogonality constraints, allowing the eigensolver itself to adapt to the classification objective while retaining the computation structure of the Lanczos procedure.

Mathematically, we establish conditions under which $\u_1$ can be accurately approximated after a finite number of Lanczos iterations. 
These conditions motivate regularization terms during unrolling that encourage accurate eigenvector estimation while optimizing end-to-end classification performance.
Experiments show that our network has comparable classification performance to state-of-the-art (SOTA) deep learning models \cite{chen2017deeplab, xie2021segformer}, while reducing parameter count by up to two orders of magnitude. 

We summarize our main contributions as follows:

\begin{itemize} 
\item We formulate binary crop classification as a signed graph spectral clustering problem and develop a lightweight signed graph spectral network that jointly learns graph structure and classification labels. 
\item We derive theoretical conditions guaranteeing accurate finite-iteration approximation of the smallest-eigenvalue eigenvector of a hybrid graph Laplacian using Lanczos iterations. 
\item We propose a Lanczos-unrolled neural architecture that relaxes classical orthogonality constraints and incorporates theory-driven regularization for end-to-end learning. 
\item We demonstrate on harmonized Landsat-Sentinel imagery that the proposed approach achieves competitive classification performance with state-of-the-art semantic segmentation models while reducing the number of trainable parameters by up to two orders of magnitude. 
\end{itemize}



\section{Related Work}
\label{sec:related}

\subsection{Crop Classification}

Crop classification from remote sensing imagery is commonly formulated as a semantic segmentation problem, where each pixel is assigned a land-cover or crop category. 
Deep learning has become the dominant paradigm for this task, with numerous convolutional and attention-based architectures proposed for dense pixel-wise prediction \cite{UPerNet, kirillov2017unified, zhao2017pyramid, li2018pyramid, long2015fully, chen2019rethinking, MAnet}. 

\textit{Fully convolutional networks} (FCNs) first established end-to-end semantic segmentation by employing  convolutional operations, enabling dense prediction directly from input images \cite{long2015fully}. 
Subsequent methods focused on incorporating multi-scale contextual information. 
PSPNet introduced pyramid pooling to aggregate features from regions of different spatial scales \cite{zhao2017pyramid}, while DeepLab employed atrous convolution and atrous spatial pyramid pooling to enlarge the receptive field \cite{chen2019rethinking}. 
PANet and UPerNet further improved dense prediction by combining multi-scale feature aggregation with attention mechanisms and feature pyramid representations \cite{li2018pyramid,UPerNet}. 
Attention-based architectures have also demonstrated strong performance in dense prediction tasks; for example, MA-Net leveraged spatial and channel attention to refine feature representations for medical image segmentation \cite{MAnet}. 

Despite their impressive performance, pure deep learning-based segmentation methods typically require a large number of trainable parameters and substantial computational resources. 
Moreover, their highly nonlinear architectures make it challenging to interpret the operational decision process. 

\subsection{Clustering on Positive Graphs}

Clustering algorithms on positive graphs often rely on graph-cut formulations. 
Given a positive weighted graph $\cG=(\cV,\cE)$, a cut partitions the vertex set into two disjoint subsets, $\cA$ and $\cV\setminus\cA$, and its cost, $\operatorname{cut}(\cA,\cV\setminus\cA)$, is the total weight of the edges connecting nodes from one subset to the other. 
The \textit{minimum-cut} problem seeks the partition with the lowest such cost, but it may produce a trivial solution in which a very small vertex subset is weakly connected with others. 

\textit{Ratio cut} reduces this tendency by normalizing the cut cost according to the sizes of the two subsets:
\begin{equation}
\operatorname{RatioCut}(\cA,\cV\setminus\cA)
=
\frac{\operatorname{cut}(\cA,\cV\setminus\cA)}{|\cA|}+\frac{\operatorname{cut}(\cA,\cV\setminus\cA)}{|\cV\setminus\cA|}.
\label{eq:ratiocut}
\end{equation}
The inverse-cardinality terms penalize unbalanced partitions, making the isolation of a small set of vertices more costly. 
A spectral relaxation of RatioCut is obtained using the eigenvector $\v_2$ associated with the second-smallest eigenvalue of the unnormalized graph Laplacian $\L=\D-\W$ \cite{von2007tutorial}. 
\textit{Normalized cut} normalizes each partition using its total connectivity to the graph, rather than only its number of vertices \cite{shi00}:
\begin{equation}
\operatorname{Ncut}(\cA,\cV\setminus\cA)
=\frac{\operatorname{cut}(\cA,\cV\setminus\cA)}{\operatorname{assoc}(\cA,\cV)}
+\frac{\operatorname{cut}(\cA,\cV\setminus\cA)}{\operatorname{assoc}(\cV\setminus\cA,\cV)},
\label{eq:ncut}
\end{equation}
where $\operatorname{assoc}(\cA,\cV)=\sum_{i\in\cA,\,j\in\cV}w_{i,j}$ denotes the total connection from $\cA$ to all vertices in the graph. 
The spectral relaxation of Ncut uses the second eigenvector of the normalized graph Laplacian $\cL$ rather than the unnormalized Laplacian $\L$.

We note that a strictly positive graph means that only pairwise \textit{similarities} are encoded as positive edge weights. 

\subsection{Clustering on Signed Graphs}

The vast majority of research in GSP has focused on positive graphs \cite{ortega18ieee}. 
In contrast, \textit{signed graphs} incorporate both positive and negative edge weights, encoding pairwise similarities and dissimilarities, respectively. 
Signed graphs were used in \cite{harary1953notion} to study the dynamics of groups from a social-psychological perspective. 
\cite{hou2003laplacian} introduced the \textit{signed Laplacian} and provided a theoretical analysis of its largest eigenvalue.

\cite{Kunegis2010} introduced the \textit{SignedRatioCut} formulation, which seeks a partition that minimizes the weights of positive edges crossing between clusters and the weights of negative edges within each cluster. 
The authors showed that the solution to a relaxation of SignedRatioCut is the eigenvector $\u_1^s$ corresponding to the smallest eigenvalue $\lambda_1$ of the signed Laplacian. Subsequently, the graph can be bi-partitioned according to the signs of $\u_1^s$'s entries.
How the underlying signed graph was constructed was not discussed.

\cite{knyazev2017spectral} compared the standard combinatorial Laplacian with the signed Laplacian and argued that the former exhibits a larger eigen-gap around its smallest eigenvalue, potentially leading to faster convergence and greater numerical stability in eigenvector computation. 
Similarly, we argue that eigenvectors of both the combinatorial and signed Laplacian carry complementary information, and thus propose a tunable convex combination of the two Laplacians as part of our clustering solution. 

\cite{mercado2019spectral} modeled positive and negative interactions using two separate adjacency matrices, $\mathbf{W}^{+}$ and $\mathbf{W}^{-}$, and combine the corresponding normalized Laplacians through a matrix power mean. 
Here, a node pair $(i,j)$ may exhibit both positive and negative interactions, represented by $W^{+}_{i,j}$ and $W^{-}_{i,j}$, respectively. 
Such a multilayer  representation is well suited to applications where attractive and repulsive relations constitute distinct sources of information. 
In contrast, our crop-classification task considers a single graph derived from learned feature representations, where pairwise relationships are modeled through similarity measures. 
Thus, a single-layer graph formulation is sufficient for our problem setting.


\cite{matz2020learning} proposed a framework for estimating a signed graph directly from an observation matrix by decoupling the optimization of edge signs and magnitudes, reducing the learning problem to a quadratic program with linear constraints. 
The learned graph is subsequently used for downstream tasks such as total variation (TV) clustering \cite{bresson2013multiclass}. 
While \cite{matz2020learning} is likewise data-driven, our signed-graph learning and clustering framework differs in three important respects. 
First, \cite{matz2020learning} learned a signed graph directly from an input data matrix, whereas our graph is constructed from learned node features $\mathbf{f}_i$ produced by a lightweight CNN.
Thus, graph construction in our framework is driven by representation learning rather than a standalone graph-learning optimization problem. 
Second, our clustering method is \textit{spectral} in nature, relying on Laplacian eigenvectors for binary classification, whereas TV clustering \cite{bresson2013multiclass} is based on the minimization of an $\ell_1$-regularized objective and does not admit the same frequency-domain interpretation. 
Finally, graph learning and clustering are decoupled in \cite{matz2020learning}, while our unrolled network jointly learns feature representations, constructs the signed graph, and performs clustering in an end-to-end trainable framework.

\cite{karaaslanli2022scsgl} proposed scSGL, a kernelized signed graph learning method for inferring gene regulatory networks from single-cell RNA-seq data. 
Similar to \cite{mercado2019spectral}, the method represents positive and negative interactions using two adjacency matrices, $\mathbf{W}^{+}$ and $\mathbf{W}^{-}$. 
However, a mutual-exclusivity constraint is imposed so that each node pair $(i,j)$ may carry either a positive or a negative interaction, but not both. 
Since our formulation adopts the same exclusivity assumption, we represent pairwise relationships using a single signed adjacency matrix and perform spectral clustering on the corresponding graph Laplacian.

\subsection{Graph Algorithm Unrolling}

\textit{Algorithm unrolling} \cite{monga21} maps a fixed number of iterations of an optimization algorithm to a sequence of neural layers, enabling selected parameters in the layers to be learned in a data-driven manner while retaining interpretability. 
A notable example is \cite{yu2023white}, where unrolling an algorithm minimizing a sparse rate reduction (SRR) objective led to an interpretable transformer-like architecture with competitive imaging performance. 
Specifically for GSP, \cite{thuc2024interpretable} showed that graph learning with normalization is a form of self-attention, resulting in a transformer-like model via graph algorithm unrolling. 
Related works further explored unrolling on positive graphs for image restoration \cite{cai2025unrolling}, balanced signed graphs for EEG classification \cite{yao2026lightweight}, and directed graphs for traffic prediction \cite{qi2025lightweight}. 
Unlike existing graph-unrolling methods, which employ graph regularization terms such as \textit{graph Laplacian regularizer} (GLR) \cite{pang17}, \textit{graph total variation} (GTV) \cite{bai19} and \textit{gradient graph Laplacian regularizer} (GGLR) \cite{chen24} to promote low-frequency solutions, our Lanczos unrolling explicitly computes a graph frequency---the first eigenvector of a hybrid graph Laplacian---for binary classification.



\begin{figure*}[t]
    \centering
    \includegraphics[width=0.8\textwidth]{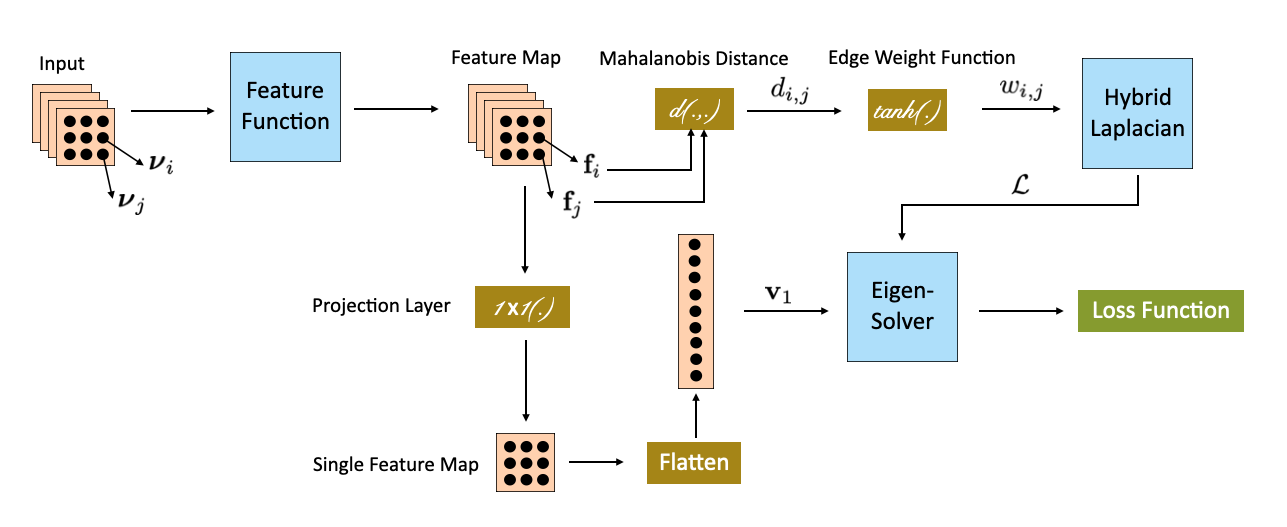}
    \captionsetup{justification=centering}
    \vspace{-0.2in}
    \caption{Illustration of overview architecture of the proposed model.}
    \label{fig:full_width}
\end{figure*}

\section{Preliminaries}
\label{sec:prelim}

\subsection{Definitions of Graph Laplacians}
\label{subsec:defn}

An undirected graph $\cG = (\cV, \cE, \W)$ is defined by a node set $\cV = \{1, \ldots, N\}$ and an edge set $\cE = \{(i,j)\}$ with $M = |\cE|$ edges. 
Each edge $(i,j) \in \cE$ has weight $w_{i,j} = W_{i,j} \in \mathbb{R}$, for \textit{adjacency matrix} $\W \in \mathbb{R}^{N \times N}$. 
Undirected edges mean $W_{i,j} = W_{j,i}$, and $\W$ is symmetric.
Diagonal \textit{degree matrix} $\D \in \mathbb{R}^{N \times N}$ has entries $D_{i,i} = \sum_j W_{i,j}$. 
The \textit{combinatorial graph Laplacian matrix} $\L \in \mathbb{R}^{N \times N}$ \cite{ortega18ieee} is defined as
\begin{align}
\L \triangleq \D - \W = \mathrm{diag}(\W \1) - \W, 
\label{eq:combL}
\end{align}
where $\1$ is the all-one vector of appropriate length, and $\mathrm{diag}(\v)$ is a diagonal matrix containing vector $\v$ as diagonal entries.
$\L$ is provably \textit{positive semi-definite} (PSD) if all edge weights are non-negative, \ie, $w_{i,j} \geq 0, \forall i,j$ \cite{cheung18}. 

For a signed graph $\cG^s$ where edge weight $w_{i,j}$ can be negative, a \textit{signed graph Laplacian} is defined in \cite{Kunegis2010} as
\begin{align}
\L^s \triangleq \D^s - \W = \mathrm{diag}(|\W|\1) - \W,
\label{eq:signedL}
\end{align}
where $|\W|$ returns the absolute value of $\W$ element-wise.


\subsection{Graph Balance}
\label{subsec:graph_balance}

A signed graph $\cG^s$ is \textit{balanced} if it contains no cycle with an odd number of negative edges.
Equivalently, by the \textit{Cartwright-Harary Theorem} (CHT) \cite{harary1953notion}, $\cG^s$ is balanced iff polarities $p_i \in \{1,-1\}$ can be assigned to the nodes such that each positive (negative) edge connects a node pair of the same (opposite) polarities;
mathematically, we write
\begin{align}
\mathrm{sign}(w_{i,j}) = p_i p_j, ~~~~~~ \forall (i,j) \in \cE .
\label{eq:graph_balance}
\end{align}

It is shown in \cite{yang22} that the first eigenvector $\u_1$ of the Laplacian $\L$ of a connected balanced signed graph has entries whose signs are the node polarities, \ie, $\mathrm{sign}(u_{1,i}) = p_i$, up to a global sign flip.
In binary node classification, we interpret each ground-truth label $c_i \in \{1,-1\}$ as the node polarity, $p_i = c_i$, so a signed graph is \textit{balanced with respect to the labels} if $\mathrm{sign}(w_{i,j}) = c_i c_j$ for all edges.
For such a graph, the entry signs of $\u_1$ recover the binary labels, up to a global sign flip.
In practice, a graph learned from data is only \textit{near-balanced}: some \textit{frustrated} edges violate $\mathrm{sign}(w_{i,j}) = c_i c_j$, so the signs of $\u_1$ only approximately match the labels.



\subsection{Gershgorin Circle Theorem}
\label{sec:GCT}

\textit{Gershgorin Circle Theorem} (GCT) \cite{varga04} states that each real eigenvalue $\lambda$ of a real symmetric matrix $\A$ must reside in at least one Gershgorin disc, \ie, $A_{i,i} - r_i \leq \lambda \leq A_{i,i} + r_i, ~~~\exists i$, where $r_i = \sum_{j \neq i} |A_{i,j}|$ is the radius of the $i$-th Gershgorin disc. A corollary is that the smallest Gershgorin disc left-end, $\lambda^-_{\min}(\A)$, is a lower-bound of the smallest eigenvalue $\lambda_{\min}(\A)$: $\lambda^-_{\min}(\A) = \min_i (A_{i,i} - r_i) \leq \lambda_{\min}(\A)$.

\section{Building a Graph-based Binary Classifier}
\label{sec:algo}

\subsection{System Overview}

We assume that a set $\cT$ of satellite images, each with pixel-label pairs $(i, c_i)$ identifying binary crop types $c_i \in \cC = \{1, -1\}$, are available for training.
Also associated with each pixel $i$ is frequency response vector $\bnu_i = [\nu_i^1; \ldots; \nu_i^P] \in \mathbb{R}^P$ from $P$ different hyperspectral frequencies.
Denote by $\e_i$ the \textit{embedding} of pixel $i$, which includes frequency response $\bnu_i$ of pixel $i$ and those of its nearest spatial neighbors.
We employ \textit{feature function} $F(\cdot)$ (to be discussed in Section\;\ref{subsec:unroll_train}) to compute feature vector $\f_i = F(\e_i) \in \mathbb{R}^K$ for each pixel $i$.

Next, non-negative \textit{feature distance} $d_{i,j}(\f_i,\f_j)$ is computed between neighboring pixels $i$ and $j$. 
Signed edge weight $w_{i,j} \in (-1,1)$ of an edge $(i,j) \in \cE$ connecting pixels $i$ and $j$ is mapped from $d_{i,j}(\f_i,\f_j)$ using a hyperbolic tangent function. 
Computed edge weights $\{w_{i,j}\}$ define a signed graph $\cG^s$; designed loss functions promote graph balance during training, resulting in a near-balanced graph. 
Finally, the first eigenvector $\u_1$ corresponding to the smallest eigenvalue of a hybrid Laplacian combining the signed graph Laplacian $\L^s$ and the combinatorial graph Laplacian $\L$ for $\cG^s$ is computed using a neural network constructed via unrolling of the Lanczos algorithm\footnote{Alternatively, extreme eigenvectors can be computed efficiently using \textit{Locally Optimal Block Preconditioned Conjugate Gradient} (LOBPCG) \cite{knyazev2001toward}. We leave unrolling of LOBPCG as future work.} \cite{lanczos1950iteration}. 
The signs of the entries in $\u_1$ are used to assign pixel labels (see Fig.~\ref{fig:full_width}).
Parameters of $F(\cdot)$ and the unrolled Lanczos-based eigensolver are learned via back-prop using training data $\cT$ by minimizing ends-to-end loss functions (to be discussed in Section\;\ref{sec:Training process}).

\subsection{Signed Graph Construction}
\label{subsec:construct_graph}

Given computed features $\f_i = F(\e_i)$ and a \textit{metric matrix} $\M \in \mathbb{R}^{K \times K},$ the non-negative \textit{Mahalanobis distance} (feature distance) $d_{i,j}$ between pixels (nodes) $i$ and $j$ is
\begin{align}
d_{i,j} = (\f_i - \f_j)^\top \M (\f_i - \f_j) ,
\end{align}
where $\M \succeq 0$ is symmetric and PSD, so that $d_{i,j} \geq 0, \forall \f_i, \f_j$.
Pairwise distances $d_{i,j}$ define the \textit{geometry} of the data: pixels $i$ and $j$ with similar features $\f_i$ and $\f_j$ are geometrically close with $d_{i,j} \approx 0$ (and thus likely in the same class), while pixels $i$ and $k$ with vastly different features $\f_i$ and $\f_l$ are geometrically far-apart with $d_{i,l} \gg 0$ (and thus likely in different classes).

Conventionally in graph-based binary classification \cite{shi00}, non-negative feature distance $d_{i,j}$ is mapped to \textit{non-negative} edge weight $w_{i,j} \in (0,1]$ via an exponential kernel:
\begin{align}
w_{i,j} = \exp (-d_{i,j}) .
\label{eq:weightExp}
\end{align}
Non-negative edge weights $\{w_{i,j}\}$ specify a \textit{positive} graph $\cG^+$, using which binary labels $c_i \in \{-1, 1\}$ are assigned for nodes $i$ via spectral algorithms such as \cite{shi00}.
However, for binary classification a large feature distance $d_{i,j}$ implies more than weak \textit{similarity} between node-pair $(i,j)$; a sufficiently large distance $d_{i,j}$ means nodes $i$ and $j$ are substantially different, and thus the node-pair should be assigned opposite labels. 
Thus, we employ a \textit{negative} edge to express pairwise \textit{dissimilarity} and promote opposite label assignment. 

\begin{figure}%
    \centering
    \subfloat{{\includegraphics[width=0.7\linewidth]{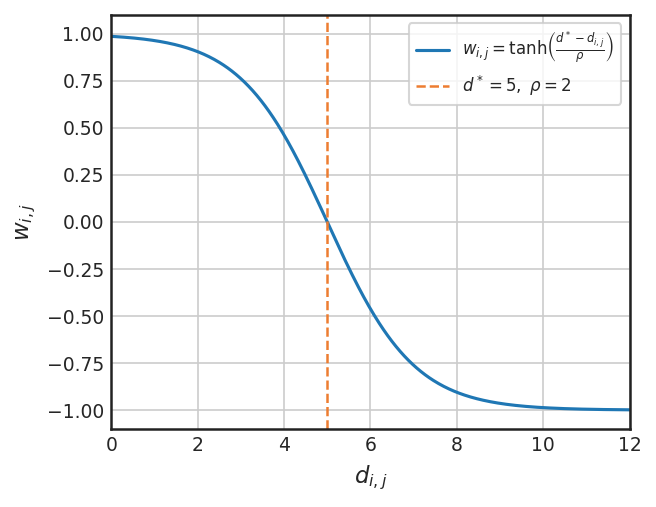}}}%
    \vspace{-0.1in}
    \caption{Illustration of the hyperbolic tangent function with $d^{*} = 5$ and $\rho = 2$ that maps non-negative feature distance $d_{i,j} \in \mathbb{R}_+$ to signed edge weight $w_{i,j} \in (-1,1)$.}%
    \label{fig:tanh}%
\end{figure}

To compute \textit{signed} edge weight $w_{i,j} \in (-1,1)$ given non-negative feature distance $d_{i,j}$, we employ the hyperbolic tangent function:
\begin{align}
w_{i,j} = \tanh \left(\frac{d^* - d_{i,j}}{\rho} \right) ,
\label{eq:tanh}
\end{align}
where $d^*, \rho \in \mathbb{R}_+$ are positive scalar parameters. 
When distance $d_{i,j} \approx 0$, edge weight $w_{i,j} \approx 1$, expressing strong pairwise similarity, meaning that node pair $(i,j)$ should have the same label.  
When distance $d_{i,j} = d^*$, edge weight $w_{i,j} = 0$, expressing pairwise ambiguity.
Finally, when distance $d_{i,j} \gg d^*$, edge weight $w_{i,j} \approx -1$, expressing strong pairwise dissimilarity, meaning that node pair $(i,j)$ should have opposite labels.
Since $\tanh(\cdot)$ is strictly monotonic and preserves distance ordering, the constructed graph $\cG^s$ remains a geometric graph.
See Fig.\;\ref{fig:tanh} for an illustration. 



\subsection{First Eigenvectors of Signed / Combinatorial Laplacians}

Given a constructed signed graph $\cG^s$, there are two common graph Laplacian matrix definitions in the GSP literature: i) combinatorial graph Laplacian $\L = \mathrm{diag}(\W\mathbf{1}) - \W$ \cite{ortega18ieee}, and ii) signed graph Laplacian $\L^s = \mathrm{diag}(|\W|\mathbf{1}) - \W$ \cite{Kunegis2010}. 
Their respective Rayleigh quotients are \begin{align} 
\x^\top \L \x &= \sum_{i<j} w_{i,j}(x_i-x_j)^2, \\ 
\x^\top \L^s \x &= \sum_{i<j} |w_{i,j}| \bigl(x_i-\mathrm{sign}(w_{i,j})x_j\bigr)^2.
\label{eq:Rayleigh_signedL}
\end{align} 
These formulations reveal that the two Laplacians interpret ``dissimilarity'' differently. 
For $\L$, a negative edge weight $w_{i,j}<0$ acts as a \textit{repulsive} force, encouraging a large inter-sample difference $|x_i-x_j|$. 
In contrast, for $\L^s$, a negative edge promotes \textit{anti-correlation}, encouraging $x_i \approx -x_j$. 

\begin{figure}%
    \centering
    \subfloat[\centering]{{\includegraphics[width=1\linewidth]{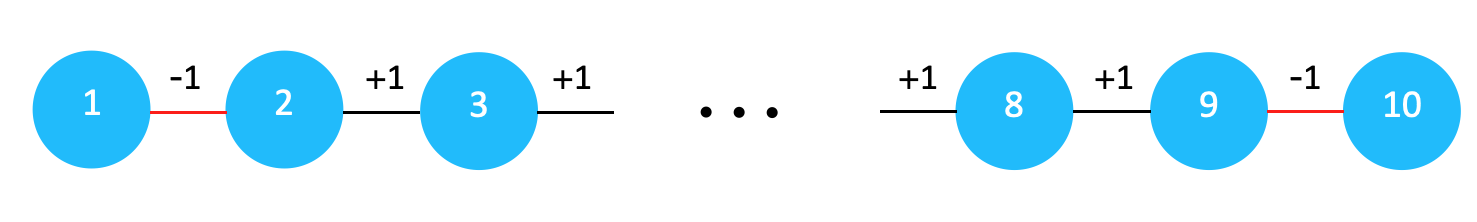}}} \\
    \vspace{-0.4cm} 
    \subfloat[\centering]{{\includegraphics[width=0.7\linewidth]{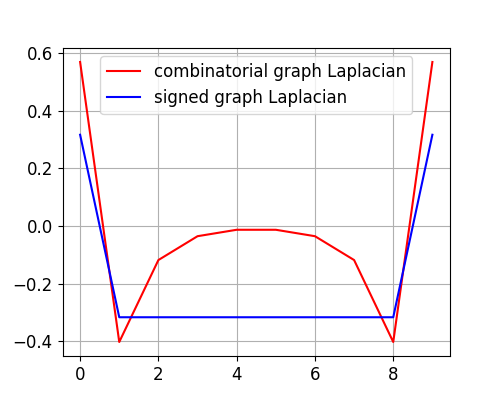}}}
    \vspace{-0.05in}
    \caption{a) Illustration of a 10-node path graph with unit edge weights $w_{i,i+1} = 1$, except for the first and last edges with weight $w_{0,1} = w_{8,9} = -1$. b) Comparison of the first eigenvectors of signed graph Laplacian $\L^s$ vs. combinatorial graph Laplacian $\L$ for the 10-node path graph.}%
    \label{fig:logistic}%
\end{figure}

Consequently, the eigenvectors associated with $\L$ and $\L^s$ exhibit markedly different behaviors. 
Fig.~\ref{fig:logistic}(b) illustrates this phenomenon for a $10$-node path graph, shown in Fig.~\ref{fig:logistic}(a), with edge weights $w_{i,j}=1$, except for the boundary edges $w_{0,1} = w_{8,9}=-1$. 
The first eigenvector of $\L^s$ is nearly constant over most nodes, reflecting the \textit{local} pairwise (dis)similarities encoded by the signed edges. 
In contrast, the first eigenvector of $\L$ exhibits a more \textit{global} effect: the entries corresponding to nodes $4$ and $5$ are more similar than those corresponding to nodes $7$ and $8$, since the former pair lies farther from the negatively weighted edges. 
This demonstrates how the repulsive influence of negative edges propagates through the graph. 
A similar effect was noted by \cite{knyazev2017spectral} in the analysis of mass-spring systems with repulsive springs, where negative edge weights induce global changes in the low-frequency vibration modes of the conventional graph Laplacian.

These observations suggest that $\L$ and $\L^s$ capture complementary information. 
To leverage both sources of information, we construct a hybrid Laplacian as a convex combination,
\begin{align} \cL = \tau \L^s + (1-\tau)\L , 
\end{align} 
where $\tau \in [0,1]$ is a parameter, and use the first eigenvector $\u_1$ of $\cL$ for binary classification.




\section{Computing First Eigenvector via Lanczos}
\label{sec:Lanczos}

Given a hybrid graph Laplacian $\cL$, we compute its first eigenvector $\u_1$ via unrolling of the \textit{Lanczos algorithm} \cite{lanczos1950iteration}. 
We first review the iterative Lanczos algorithm.

\subsection{Review of the Lanczos Algorithm}
\label{subsec:Lanczos_review}

Given a unit-norm initial vector $\v_1$ and a real symmetric matrix $\cL$, the $m$-th \textit{Krylov space} is defined as
\begin{align}
\cK_m(\cL,\v_1) \triangleq \mathrm{span}\{\v_1, \cL \v_1, \ldots, \cL^{m-1} \v_1\} .
\end{align}
The Lanczos algorithm iteratively computes $m$ orthonormal \textit{Lanczos vectors} $\V_m = [\v_1, \ldots, \v_m]$ that span this Krylov space $\cK_m(\cL,\v_1)$.
Specifically, in matrix form,  $\cL$ and $\V_m$ are related by

\vspace{-0.1in}
\begin{footnotesize}
\begin{align}
\V_m^\top \cL \V_m &= \T_m = \left[ \begin{array}{cccccc}
\alpha_1 & \beta_1 & & & & 0 \\
\beta_1 & \alpha_2 & \beta_2 & & & \\
 & \beta_2 & \alpha_3 & \ddots & & \\
& & \ddots & \ddots & \beta_{m-2} & \\
& & & \beta_{m-2} & \alpha_{m-1} & \beta_{m-1} \\
0 & & & & \beta_{m-1} & \alpha_m
\end{array}
\right] ,
\label{eq:Lanczos_matrix}
\end{align}
\end{footnotesize}\noindent
where $\T_m$ is an $m \times m$ symmetric tridiagonal matrix. 

Another interpretation of \eqref{eq:Lanczos_matrix} is that $\T_m$ is the \textit{Rayleigh-Ritz projection} of $\cL$ onto the Krylov space $\cK_m(\cL,\v_1)$. 
Specifically, if $(\theta_{\min},\y_1)$ denotes the eigen-pair of $\T_m$ corresponding to the smallest eigenvalue, \ie, $\T_m \y_1 = \theta_{\min} \y_1$, and we define the corresponding \textit{Ritz vector} as $\x_1 \triangleq \V_m \y_1$, then the residual satisfies the \textit{Galerkin orthogonality condition} \cite{saad2011numerical}
\begin{align}
\V_m^\top (\cL \x_1 - \theta_{\min} \x_1) = \0, 
\label{eq:Galerkin}
\end{align}
 or equivalently, 
\begin{align}
\cL \x_1 - \theta_{\min} \x_1 \perp \cK_m(\cL,\v_1). 
\end{align}
This interpretation shows that the Lanczos algorithm efficiently computes an approximation, the Ritz pair $(\theta_{\min},\x_1)$, to the smallest eigen-pair of $\cL$.

\subsection{Conditions of $\T_m$ for First Eigenvector Approximation}

We examine conditions under which the Lanczos tridiagonal matrix $\T_m$, characterized by diagonal entries $\alpha_i$ and off-diagonal entries $\beta_i$ as shown in \eqref{eq:Lanczos_matrix}, yields an accurate approximation of the first eigenvector $\u_1$. 
The $i$-th Lanczos vector $\v_i$ belongs to the Krylov subspace 
\begin{align} \cK_i(\cL,\v_1) = \mathrm{span} (\v_1,\cL\v_1,\ldots,\cL^{i-1}\v_1 ). 
\end{align} 
$\beta_i$ is the residual coefficient generated when orthogonalizing $\cL\v_i$ against the current Lanczos basis. 
Consequently, $|\beta_i|$ measures the strength of the newly generated direction that expands the Krylov subspace. 
Large values of $|\beta_i|$ indicate that the Krylov subspace continues to acquire substantial new information from repeated applications of $\cL$. 

On the other hand, the diagonal coefficient $\alpha_i = \v_i^\top \cL \v_i$ is the \textit{Rayleigh quotient} of $\v_i$. 
Writing 
\begin{align} \v_i = \sum_{j=1}^{N} q_j \u_j, 
\end{align} 
where $(\lambda_j,\u_j)$'s are the eigen-pairs of $\cL$, we have \begin{align} \alpha_i = \sum_{j=1}^{N} \lambda_j |q_j|^2. \end{align} 
Thus, $\alpha_i$ represents the average eigenvalue, or frequency, of $\cL$ weighted by the spectral energy distribution of $\v_i$. 
Generally, smaller values of $\alpha_i$ indicate stronger spectral concentration of $\v_i$ in the low frequencies. 

It follows that a ``good'' set of $m$ Lanczos iterates, in terms of estimating the first eigenvector $\u_1$ of $\cL$, is one with small $\alpha_i$'s and large $|\beta_i|$'s.
Formally, we first establish that imposing a uniform upper bound on $\alpha_i$ for all Lanczos vectors necessarily limits the number of orthogonal Lanczos vectors that can satisfy such a bound.
We develop the following theory for a real symmetric matrix $\cL$ with eigen-gap $\lambda_2 - \lambda_1 > 0$; hybrid Laplacian of a \textit{connected} signed graph is a special case.

\begin{lemma}[Low-frequency Concentration of Lanczos Vectors]
Let $\cL \in \mathbb{R}^{N\times N}$ be a real symmetric matrix with eigenvalues 
\begin{align}
\lambda_1 < \lambda_2 \leq \cdots \leq \lambda_N . 
\end{align} 
Let $\{ \v_i \}_{i=1}^m$ be the Lanczos basis vectors generated by $\cL$ and a unit-norm initial vector $\v_1$. 
Suppose that 
\begin{align}
\alpha_i = \v_i^\top \cL \v_i \leq \lambda_2 - \epsilon, \qquad i=1,\ldots,m,
\end{align}
for some $\epsilon>0$. 
Then 
\begin{align}
m \leq \frac{\lambda_2 - \lambda_1}{\epsilon}.
\end{align}
\label{lemma:alpha_i}
\end{lemma}\noindent
See Appendix\;\ref{append:proof_lemma1} for a proof.

Next, define $\T_m^o$ as the matrix obtained from the Rayleigh-Ritz projection $\T_m$ by setting all diagonal entries to zero. 
We establish that uniformly lower-bounding $|\beta_i|$ forces the smallest eigenvalue $\theta_1(\T_m^o)$ to decrease, and derive an explicit upper bound for it.

\begin{lemma}[Monotonicity of the Smallest Eigenvalue] 
Let 
\begin{align}
\T^o_m = \begin{bmatrix} 0 & \beta_1 & 0 & \cdots & 0\\ \beta_1 & 0 & \beta_2 & \ddots & \vdots\\ 0 & \beta_2 & 0 & \ddots & 0\\ \vdots & \ddots & \ddots & \ddots & \beta_{m-1}\\ 0 & \cdots & 0 & \beta_{m-1} & 0 \end{bmatrix}    
\end{align}
be a real symmetric tridiagonal matrix satisfying
\begin{align}
|\beta_i| \ge \beta_{\min} > 0, \qquad i=1,\ldots,m-1.    
\end{align}
Define the constant-off-diagonal tridiagonal matrix
\begin{align}
\T_{\min}= \begin{bmatrix} 0 & \beta_{\min} & 0 & \cdots & 0\\ \beta_{\min} & 0 & \beta_{\min} & \ddots & \vdots\\ 0 & \beta_{\min} & 0 & \ddots & 0\\ \vdots & \ddots & \ddots & \ddots & \beta_{\min}\\ 0 & \cdots & 0 & \beta_{\min} & 0 \end{bmatrix}.    
\end{align}
Then
\begin{align}
\theta_1(\T^o_m) \le \theta_1(\T_{\min}) = -2\beta_{\min}\cos\!\left(\frac{\pi}{m+1}\right),    
\end{align}
where $\theta_1(\cdot)$ denotes the smallest eigenvalue.
\label{lemma:beta_i}
\end{lemma}\noindent 
See Appendix\;\ref{append:proof_lemma2} for a proof.

We now state our main result, which relates the Lanczos coefficients $\alpha_i$ and $\beta_i$ to the low-frequency coverage of the Krylov subspace, the smallest Ritz value, and the alignment of the Ritz vector $\x_1$ with the first eigenvector $\u_1$.

\begin{theorem}[Low-Frequency Coverage and Ritz-Pair Approximation Bound] 
Let $\cL \in \mathbb{R}^{N\times N}$ be a real symmetric matrix with eigenvalues 
\begin{align}
\lambda_1 < \lambda_2 \le \cdots \le \lambda_N.    
\end{align}
Let $\V_m =[\v_1,\ldots,\v_m]$ be the Lanczos basis, and let 
\begin{align}
\T_m = \begin{bmatrix} \alpha_1 & \beta_1 \\ \beta_1 & \alpha_2 & \beta_2 \\ & \beta_2 & \ddots & \ddots \\ && \ddots & \alpha_{m-1} & \beta_{m-1}\\ &&& \beta_{m-1} & \alpha_m \end{bmatrix} =\V_m^\top \cL \V_m    
\end{align}
be the associated Lanczos tridiagonal matrix. 
Suppose that 
\begin{align}
\alpha_i \le \lambda_2-\epsilon, \qquad i=1,\ldots,m,     
\end{align}
for some $\epsilon>0$, and 
\begin{align}
|\beta_i|\ge \beta_{\min}>0, \qquad i=1,\ldots,m-1.    
\end{align}
Then: 
\begin{enumerate}
\item Let $\cK_m=\mathrm{span}\{\v_1,\ldots,\v_m\}$. 
If $\u_1$ denotes the eigenvector of $\cL$ corresponding to $\lambda_1$, then 
\begin{align}
\sin \angle(\u_1,\cK_m) \le \sqrt{1-\frac{m\epsilon}{\lambda_2 - \lambda_1}}.    
\end{align}
\item Let $\theta_1(\T_m)$ denote the smallest eigenvalue of $\T_m$. 
Then 
\begin{align}
\lambda_1(\cL) \le \theta_1(\T_m) \le (\lambda_2-\epsilon) - 2\beta_{\min} \cos\!\left(\frac{\pi}{m+1}\right).    
\end{align}
\item Let $\x_1=\V_m\y_1$ be the Ritz vector associated with $\theta_1(\T_m)$. 
Then
\begin{align}
|\u_1^\top \x_1|^2 \ge \frac{\epsilon + 2\beta_{\min}\cos \left(\frac{\pi}{m+1}\right)}{\lambda_2-\lambda_1},
\end{align}
and hence $\sin^2\angle(\u_1,\x_1)\le 1-\frac{\epsilon+2\beta_{\min}\cos(\pi/(m+1))}{\lambda_2-\lambda_1}$.
\end{enumerate}
Therefore, smaller diagonal coefficients $\alpha_i$ and larger off-diagonal coefficients $|\beta_i|$ tighten the lower bound on the alignment between the computed Ritz vector $\x_1$ and the desired first eigenvector $\u_1$.
\label{thm:alpha_beta}
\end{theorem}
\noindent
See Appendix\;\ref{append:proof_thm} for a proof.

Note that Theorem~\ref{thm:alpha_beta} assumes the exact Lanczos recurrence, where $\V_m$ is orthonormal and $\T_m=\V_m^\top\cL\V_m$. 
Thus, it serves as a design guideline, not a guarantee, for the relaxed recurrence developed next. 

\section{Unrolling the Lanczos Algorithm}

Since a signed graph $\cG^s$ constructed via learned feature distances $d_{i,j}$ and $\tanh$ edge weight mapping (Section\;\ref{subsec:construct_graph}) is typically only near-balanced, its first eigenvector $\u_1$ only approximately recovers the node polarities, and the classical Lanczos algorithm (Section\;\ref{subsec:Lanczos_review}), which targets $\u_1$, inherits this limitation. 
We hypothesize that tuning the output with the classification loss can compensate. 
To this end, we relax the orthogonality condition in the Lanczos recurrence with learnable parameters $\gamma^{(1)}_j,\gamma^{(2)}_j$, $j=1,\ldots,m$. 
This reshapes the tridiagonal matrix $\T_m$, and thus its first eigenvector $\y_1$ and the Lanczos basis $\V_m$, so that $\x_1=\V_m\y_1$ remains in $\cK_m(\cL,\v_1)$ but is now tuned for binary classification.

\subsection{Relaxing Lanczos' Orthogonality Constraint}

We first write Lanczos in iterative algebraic form:
\begin{align}
    \w_j &= \cL \v_j, ~~~ \alpha_j = \v_j^\top \w_j
    \label{eq:Lanczos_alpha}\\
    \r_j &= \w_j - \alpha_j \v_j - \beta_{j-1} \v_{j-1} 
    \label{eq:Lanczos_iterative} \\
    \beta_j &= \|\r_j\|, ~~~ \v_{j+1} = \r_j/\beta_j
    \label{eq:Lanczos_beta}
\end{align}
where \eqref{eq:Lanczos_iterative} can be interpreted as an orthogonalization step, so that $\r_j$ (and thus subsequent normalized $\v_{j+1}$) is orthogonal to previously computed $\v_j, \v_{j-1}, \ldots, \v_1$, akin to the \textit{Gram-Schmidt procedure} \cite{leon2013gram}.
This is done so that the eventual Rayleigh-Ritz projection of $\cL$ on $\cK_m(\cL,\v_1)$ satisfies the Galerkin orthogonality condition \eqref{eq:Galerkin}, resulting in a good first eigen-pair approximation $(\theta_{\min},\x_1)$. 

Instead of \eqref{eq:Lanczos_iterative}, we relax the orthogonality condition and introduce learnable parameters $\gamma^{(1)}_j$ and $\gamma^{(2)}_j$ at each step $j$:
\begin{align}
\r_j &= \w_j - \gamma^{(1)}_{j} \alpha_j \v_j - \gamma^{(2)}_{j} \beta_{j-1} \v_{j-1} .
\label{eq:Lanczos_iterative2}
\end{align}
Unlike the classical Lanczos recurrence, the resulting vectors are generally not mutually orthogonal. 
Consequently, the associated tridiagonal matrix $\T_m$ is no longer symmetric in the general case when there exists a $\gamma^{(2)}_i \neq 1$, 

\vspace{0.1in}
\begin{small}
\begin{align}
\T_m = \left[ 
\begin{array}{ccccc}
\gamma^{(1)}_1 \alpha_1 & \gamma^{(2)}_2 \beta_1 & & & 0 \\
\beta_1 & \gamma^{(1)}_2 \alpha_2 & \gamma^{(2)}_3 \beta_2 & & \\
& \beta_2 & \ddots & \ddots & \\
& & \ddots & \gamma^{(1)}_{m-1} \alpha_{m-1} & \gamma^{(2)}_m \beta_{m-1} \\
0 & & & \beta_{m-1} & \gamma^{(1)}_m \alpha_m
\end{array}
\right]
\end{align}
\end{small}\noindent
and the relation $\T_m=\V_m^\top\cL\V_m$ no longer holds. 
Thus, $\T_m$ cannot be interpreted as the Rayleigh--Ritz projection of $\cL$ onto the subspace spanned by $\V_m$, and the resulting Ritz vectors do not satisfy the Galerkin orthogonality condition. 
Nevertheless, because each iterate is generated from $\cL\v_j$, $\v_j$, and $\v_{j-1}$, the vectors $\{\v_1,\ldots,\v_m\}$ continue to span the same Krylov subspace $\cK_m(\cL,\v_1)$, provided that no breakdown occurs.
The degrees of freedom introduced by the relaxation \eqref{eq:Lanczos_iterative2} allow the tridiagonal matrix $\T_m$ to adapt to the training data for downstream binary classification, unlike the classical Lanczos recurrence that minimizes the Rayleigh quotient. 

One caveat is that a general nonsymmetric tridiagonal matrix may possess complex eigenvalues. 
To ensure that the spectrum of $\T_m$ remains real, we impose the constraint  $\gamma^{(2)}_j > 0, \forall j$. 
Under this condition, $\T_m$ is diagonally similar to a symmetric tridiagonal matrix and thus has real eigenvalues.
We formalize this in the following lemma and corollary.

\begin{lemma} 
Suppose the real tridiagonal matrix $\T_m$ has diagonal entries $a_i$, upper off-diagonal entries $u_i$, and lower off-diagonal entries $\ell_i$, satisfying $\ell_j u_{j-1}>0$ for $j=2,\ldots,m$. Its form is
\begin{align}
\T_m = \begin{bmatrix} a_1 & u_1 & & & 0 \\ \ell_2 & a_2 & u_2 & & \\ & \ell_3 & \ddots & \ddots & \\ & & \ddots & a_{m-1} & u_{m-1} \\ 0 & & & \ell_m & a_m \end{bmatrix},
\end{align} 
Then, $\T_m$ is diagonal similar to a symmetric tridiagonal matrix. Consequently, all eigenvalues of $\T_m$ are real. 
\label{lemma:real_evalue}
\end{lemma}\noindent
See Appendix\;\ref{append:proof_lemma3} for a proof.

\begin{corollary} 
For $\T_m$ constructed from \eqref{eq:Lanczos_iterative2}, $\ell_j=\beta_{j-1}$ and $u_{j-1}=\gamma^{(2)}_j\beta_{j-1}$, so $\ell_j u_{j-1}=\gamma^{(2)}_j\beta_{j-1}^2$. 
Hence, if $\gamma^{(2)}_j>0$ for all $j$ and $\beta_{j-1}\neq 0$, all eigenvalues of $\T_m$ are real by Lemma~\ref{lemma:real_evalue}. 
\end{corollary}


\subsection{Shifted Inverse Iteration Method}

Given a computed tridiagonal matrix $\T_m \in \mathbb{R}^{m \times m}$, we compute the first eigenvector $\y_1$ corresponding to the smallest real eigenvalue $\theta_{\min}$ via the \textit{Shifted Inverse Iteration Method} (SIIM) \cite{saad2011numerical}. 
Specifically, if $\T_m \y_1 = \theta_{\min} \y_1$, then
\begin{align}
 (\T_m - \sigma \I)^{-1} \y_1 = \frac{1}{\theta_{\min} - \sigma}\y_1 .
\end{align}

Consequently, assuming that $\sigma < \theta_{\min}$, $\y_1$ is the eigenvector to matrix $(\T_m - \sigma \I)^{-1}$ corresponding to the \textit{largest} eigenvalue $\frac{1}{\theta_{\min} - \sigma}$. 
Standard power iteration can then be applied to this inverted system to rapidly isolate the desired $\y_1$.
Specifically, at each iteration $t$, given previous solution estimate $\y^t$, we solve a linear system for $\z^{t+1}$ then normalize:
\begin{align}
(\T_m - \sigma \I) \z^{t+1} = \y^t, 
~~~~~~ \y^{t+1} = \frac{\z^{t+1}}{\|\z^{t+1}\|} .
\label{eq:linSys_power}
\end{align}
Because $\T_m - \sigma \I$ is tridiagonal, linear system \eqref{eq:linSys_power} can be solved efficiently via the \textit{Thomas Algorithm}---a Gaussian elimination procedure for tridiagonal matrices---in $\cO(m)$ \cite{thomas1949elliptic}. 

The one remaining problem is to choose $\sigma$ such that it is near $\theta_{\min}$ while remaining $\sigma < \theta_{\min}$. 
Leveraging GCT discussed in Section\;\ref{sec:GCT}, We can define $\sigma$ as $\sigma = \lambda^-_{\min}(\T_m) - \varepsilon_{\mathrm{shift}}$, where $\varepsilon_{\mathrm{shift}}>0$ ensures that the shift remains strictly below the smallest eigenvalue.

\subsection{Algorithm Unrolling \& Training Objectives}
\label{subsec:unroll_train}


\subsubsection{Unrolling the Lanczos Algorithm}

We unroll a fixed number of iterations of the relaxed Lanczos algorithm, given by \eqref{eq:Lanczos_alpha}, \eqref{eq:Lanczos_iterative2}, and \eqref{eq:Lanczos_beta}, into neural layers to construct a feed-forward network for parameter optimization. 
Parameters $\gamma^{(1)}_{j}$ and $\gamma^{(2)}_{j}$ are initialized to $1$, corresponding to classical Lanczos, and are subsequently tuned.
Motivated by Theorem~\ref{thm:alpha_beta}, we use \textit{projected gradient descent} (PGD) to constrain the parameters so that $\gamma^{(1)}_j \leq 1$ and $\gamma^{(2)}_j \geq 1$, thereby promoting small diagonal and large-magnitude off-diagonal coefficients in $\T_m$.
We also unroll a fixed number of iterations of SIIM, given by \eqref{eq:linSys_power}, to extract the first eigenvector $\y_1$ of $\T_m-\sigma\I$.

\begin{figure}%
    \centering
\subfloat{{\includegraphics[width=9cm]{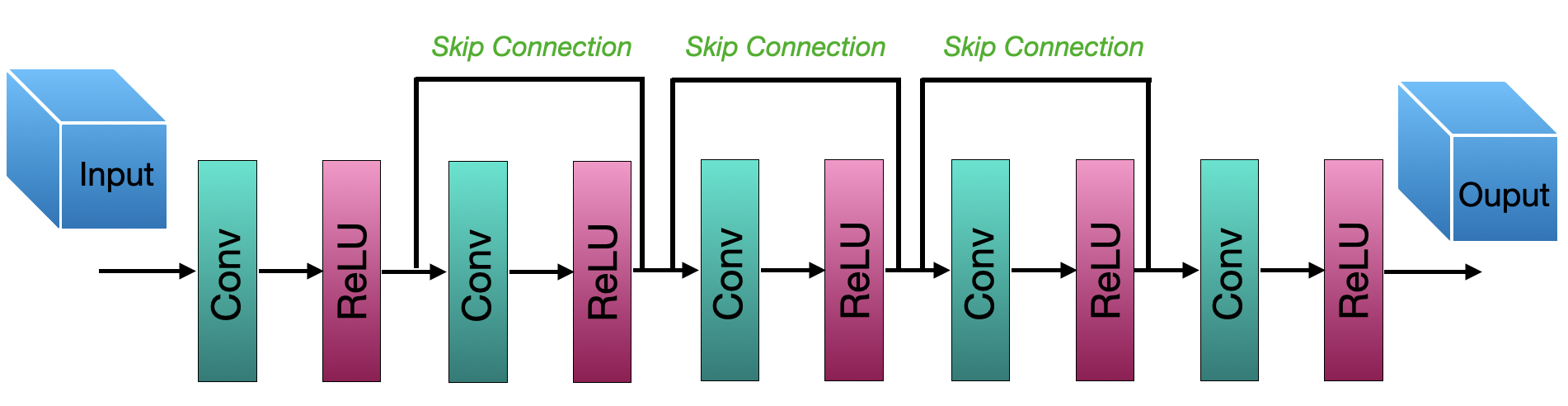} }}%
    \vspace{-0.1in}
    \caption{Architecture of the lightweight CNN, which maps an $18 \times H \times W$ input tensor to an $18 \times H \times W$ output tensor. Five of the ten convolutional layers are shown.}%
    \label{fig:CNN}%
\end{figure}

\vspace{0.05in}
\noindent
\textbf{Feature Learning:} The input to feature function $F(\cdot)$ is a concatenation of the six spectral bands (Blue, Green, Red, NIR, SWIR-1, and SWIR-2) across three time instants, yielding an input tensor of size \(18 \times H \times W\). 
We implement $F(\cdot)$ as a lightweight CNN comprising 10 convolutional layers, each followed by a ReLU activation. 
Each convolutional layer uses a \(9 \times 9\) kernel, with padding applied to preserve the spatial dimensions throughout the network. 
The first layer maps $18$ input channels to $36$ output channels, the next $8$ layers maintain $36$ channels, and the final layer maps $36$ channels to $18$. 
Skip connections are incorporated to facilitate training and improve convergence. Fig.~\ref{fig:CNN} shows the network architecture.

To accelerate convergence, we leverage the learned representation by $F(\cdot)$ to construct the starting vector \(\mathbf{v}_1 \in \mathbb{R}^{HW}\). Specifically, a point-wise \(1 \times 1\) convolutional projection layer maps feature vector \(\f_i\) at each pixel to a scalar, yielding an \(H\times W\) scalar-valued map that is vectorized to obtain \(\mathbf{v}_1\).

\subsubsection{Training process}
\label{sec:Training process}
We employ a two-stage training process. 
The first stage learns a near-balanced signed graph from data, while the second refines the graph and optimizes the unrolled Lanczos module for binary classification.

\vspace{0.05in}
\noindent\textbf{Contrastive Learning.} 
The first objective, given a ground-truth satellite image with label pairs $(i,c_i)$ for each pixel $i$, is an edge-weighted contrastive learning objective:
\begin{align}
\min_{\M, F} \!\!\! \sum_{(i,j) \in \cE | c_i = c_j} \!\!\!\!\!\!\! (2-w_{i,j}) d_{i,j} + \!\!\!\!\!\! \sum_{(i,l) \in \cE | c_i = -c_l} \!\!\!\!\!\!\!\!\!\!\! (2 +w_{i,l}) \left[ \delta - d_{i,l} \right]_+
\label{eq:contrastive2}
\end{align}
Similar to conventional contrastive learning objective \cite{hadsell2006dimensionality, schroff2015facenet, radford2021learning}, \eqref{eq:contrastive2} seeks to minimize distances between same-label pairs $(i,j)$, while maximizing distances between different-label pairs $(i,l)$ up to a margin $\delta > 0$. 
A distance $d_{i,l} = \delta$ is deemed sufficiently large to inform the classifier that node pair $(i,l)$ ought to be assigned opposite labels. 

Since $w_{i,j}$ is a decreasing function of $d_{i,j}$ via \eqref{eq:tanh}, \eqref{eq:contrastive2} naturally promotes \textit{balance} in the learned signed graph with respect to labels $\c$: same-label edges are driven towards positive weights $w_{i,j}\to+1$ and different-label edges towards negative weights $w_{i,l}\to-1$. 
In practice, some frustrated edges remain, so the learned signed graph $\cG^s$ is only \textit{near-balanced}.


\vspace{0.05in}
\noindent\textbf{End-to-End Learning.} 
In the second training phase, we jointly optimize the graph-learning and spectral-inference modules in an end-to-end manner using a two-term objective comprising a label-alignment loss and a spectral-regularization loss. The label-alignment loss is designed to encourage the estimated eigenvector to align with the ground-truth label vector while remaining invariant to a global sign reversal.

Denote by $\c$ the ground-truth binary label vector, where $c_i \in \{1,-1\}$. A fully connected signed adjacency matrix can be constructed as $\W^g=\c\c^\top-\I_N$, and the corresponding signed graph Laplacian is
\begin{align}
\L^g &= \mathrm{diag}(|\W^g|\1)-\W^g 
\\
&= (N-1)\I_N-\W^g = N\I_N-\c\c^\top.
\end{align}
The quadratic variation of $\u_1$ defines the label-alignment loss:
\begin{align}
\mathcal{J}_{\mathrm{align}}
&= \u_1^\top \L^g \u_1 
= N\|\u_1\|_2^2 - |\c^\top\u_1|^2.
\end{align}
Since $\|\u_1\|_2=1$, minimizing $\mathcal{J}_{\mathrm{align}}$ is equivalent to maximizing the squared alignment $|\c^\top\u_1|^2$, which is invariant to a global sign reversal of $\u_1$.

Motivated by Theorem~\ref{thm:alpha_beta}, which establishes that smaller Lanczos diagonal coefficients $\alpha_i$ and larger-magnitude off-diagonal coefficients $|\beta_i|$ favor a more accurate approximation of the first eigenvector, we introduce the following spectral-regularization loss:
\begin{equation}
\label{eq:alpha_beta_regularizer}
\mathcal{J}_{\mathrm{reg}}
=\frac{1}{m}\sum_{i=1}^{m}\alpha_i
+\frac{1}{m-1}\sum_{i=1}^{m-1}\frac{1}{|\beta_i|+\varepsilon_{\mathrm{div}}},
\end{equation}
where $m\geq2$ is the number of unrolled Lanczos iterations and $\varepsilon_{\mathrm{div}}>0$ prevents division by zero. 
The total training objective is a convex combination with parameter $\eta \in \mathbb{R}_+$:
\begin{equation}
\label{eq:total_loss}
\mathcal{J}_{\mathrm{total}}
=\eta\,\mathcal{J}_{\mathrm{align}}+(1-\eta)\,\mathcal{J}_{\mathrm{reg}},
\qquad 0\leq\eta\leq1.
\end{equation}


\section{Experiments}
\label{sec:exp}

\vspace{-0.05in}
\subsection{Experimental Setup}
\label{subsec:exp_setup}

\noindent\textbf{Data:} We construct a satellite image dataset built from the Harmonized Landsat–Sentinel (HLS) product \cite{claverie2018harmonized}, covering Illinois, Iowa, Nebraska, and North Dakota in the US. 
It consists of 1,500 satellite images with a spatial resolution of 
$224 \times 224$. Each image is formed by stacking six spectral bands—Blue, Green, Red, NIR, SWIR-1, and SWIR-2—across three time instants. 
The selected images correspond to the months of June, July, and August, which represent the primary growing season.

We focus on four crop types: Corn, Soybeans, Spring Wheat, and Grassland/Pasture. 
For each binary classification problem (\eg, corn vs. non-corn), we pre-filter to ensure that the target crop is sufficiently represented but does not dominate an image. 
Specifically, we retain only those images where the crop of interest covers at least 10\% and at most 90\% of the area. 
Following common practice, we partition the dataset into training, validation, and test sets using a 70/15/15\% split.

\vspace{0.05in}
\noindent\textbf{Contrastive Learning:} During contrastive training, we use a batch size of 32 and randomly sample 20,000 pixel pairs per batch, consisting of 10,000 same-label pairs and 10,000 different-label pairs. 
Hyperparameters $\delta$, $\rho$, and $d^{*}$ are set to 5, 1, and 1, respectively. 
To reduce computation cost  during inference, each $224 \times 224$ image is divided into four $112 \times 112$ sub-images--each represented by a graph with $112 \times 112 = 12{,}544$ nodes. 
Each node is connected to all nodes within a $30 \times 30$ neighborhood centered at that node. 
The model is trained for 100 epochs during this phase, with an initial learning rate of $10^{-3}$ that is reduced to $10^{-4}$ after 50 epochs.

\vspace{0.05in}
\noindent\textbf{End-to-End Learning.} In the end-to-end training phase, we use a batch size of 128 and train for 5 epochs with a learning rate of $10^{-5}$ for all crops. 
Here, parameters $d^*$, $\rho$, and $\tau$ are treated as learnable parameters and jointly optimized with the rest of the network. 
Parameters $\gamma_j^{(1)}$ and $\gamma_j^{(2)}$ of the unrolled Lanczos module are initialized to $1$ at the beginning of this phase. To determine the optimal numbers of unrolled layers for Lanczos  and SIIM, as well as the value of $\tau$, we perform a grid search over $\{10,15,20\}$, $\{2,3,4,5\}$, and $\{0.0,0.1,0.2,\ldots,0.9,1.0\}$, respectively, for each crop. 
The selected configurations are summarized in Table~\ref{tab:end_to_end_config}. 
For the $\mathcal{J}_{\mathrm{align}}$ loss term, we connect each node to its $50\times50$ neighborhood centered at that node for all crops. $\eta$ is set to $0.9$ for all crops. During inference, we resolve the sign ambiguity of the first eigenvector using feature statistics computed from the training set.

\vspace{0.05in}
\noindent\textbf{Comparison Schemes.} We select PSPNet~\cite{zhao2017pyramid}, PAN~\cite{li2018pyramid}, MA-Net~\cite{MAnet}, FPN~\cite{kirillov2017unified}, FCN~\cite{long2015fully}, DeepLabV3~\cite{chen2019rethinking}, SegFormer~\cite{xie2021segformer}, UPerNet~\cite{UPerNet}, Ncut~\cite{shi00}, and a LightweightCNN+Linear baseline for comparison. 
LightweightCNN+Linear employs the same CNN architecture used in our model, followed by a $1\times1$ point-wise convolution for pixel-wise classification. 
Except for Ncut, all models are trained for 100 epochs with a batch size of 16 and an initial learning rate of $10^{-3}$. 
For Ncut, we use the graph learning module pretrained during the contrastive learning phase.

\begin{table*}[h]
\centering
\caption{Selected numbers of layers for the unrolled Lanczos and SIIM and selected parameter $\tau$ for each crop.}
\label{tab:end_to_end_config}
\begin{small}
\begin{tabular}{lcccc}
\toprule
Configuration & Corn & Soybeans & Spring Wheat & Grassland/Pasture \\
\midrule
Unrolled Lanczos Layers
& 20
& 20
& 15
& 20 \\

Unrolled SIIM Layers 
& 3
& 3
& 3
& 3 \\

Blending parameter ($\tau$)
& 0.9
& 1.0
& 0.0
& 0.1 \\

\bottomrule
\end{tabular}
\end{small}
\end{table*}

\begin{table*}[h]
    \centering
    \caption{Results (F1 Score / Accuracy) across crops with parameter counts. In each column, the top / second / third performing scheme is highlighted in boldface / underline / italic font.}
    \label{tab:comparison}
    \begin{small}
    \begin{tabular}{lccccc}
        \toprule
        Method & Params & Corn & Soybeans & Spring Wheat & Grassland/Pasture \\
        \midrule
        Combinatorial Laplacian & \green{945,450}    & 0.9149 / 0.9444 & 0.8731 / 0.9301 & 0.7905 / 0.9037 & 0.7597 / 0.8270 \\
        Signed Laplacian        & \green{945,450}    & 0.9161 / 0.9452 & 0.8765 / 0.9321 & 0.7917 / 0.9042 & \textit{0.7607} / 0.8277 \\
        Hybrid Laplacian \cite{eftekhar2026crop} & \green{945,450}  & 0.9191 / 0.9456 & 0.8775 / 0.9338 & \textit{0.8169} / 0.9142 & \underline{0.7613} / 0.8283 \\
        \textbf{SLUNet} & \green{$\leq 945{,}512$}  & \underline{0.9193} / \underline{0.9461} & \textbf{0.8871} / 0.9376 & \textbf{0.8313} / 0.9179 & \textbf{0.7736} / 0.8394 \\
        PSPNet \cite{zhao2017pyramid}                  & 24,352,642   & 0.8900 / 0.9275 & 0.8531 / 0.9261 & 0.7929 / 0.9127 & 0.7071 / 0.8225 \\
        PAN \cite{li2018pyramid}                      & 29,040,265   & 0.9100 / 0.9408 & 0.8751 / 0.9379 & 0.8110 / 0.9175 & 0.7330 / 0.8386 \\
        MA-Net \cite{MAnet}                    & 147,487,178  & \textbf{0.9203} / \textbf{0.9483} & 0.8822 / \underline{0.9415} & 0.8071 / 0.9166 & 0.7415 / \textbf{0.8466} \\
        FPN \cite{kirillov2017unified}                      & 26,163,202   & 0.9127 / 0.9428 & 0.8825 / \textit{0.9411} & 0.8131 / \underline{0.9205} & 0.7459 / \textit{0.8419} \\
        FCN \cite{long2015fully}                      & 32,994,306   & 0.8937 / 0.9304 & 0.8527 / 0.9254 & 0.7949 / 0.9127 & 0.7103 / 0.8318 \\
        DeepLabV3 \cite{chen2019rethinking}                & 39,681,026   & 0.8932 / 0.9280 & 0.8578 / 0.9275 & 0.8054 / 0.9143 & 0.7078 / 0.8294 \\
        SegFormer \cite{xie2021segformer}               & 3,738,949    & 0.9130 / 0.9427 & \textit{0.8833} / \textit{0.9411} & \underline{0.8183} / \textbf{0.9218} & 0.7387 / 0.8377 \\
        UPerNet \cite{UPerNet}                 & 37,331,970   & \textit{0.9172} / \underline{0.9461} & \underline{0.8835} / \textbf{0.9429} & 0.8159 / \textit{0.9202} & 0.7412 / \underline{0.8465} \\
        Ncut \cite{shi00}                     & \green{945,450}   & 0.4615 / 0.5425 & 0.4128 / 0.5431 & 0.3822 / 0.5505 & 0.4295 / 0.5498 \\
        LightweightCNN+linear  & \green{945,450}   & 0.9100 / 0.9414 & 0.8453 / 0.9220 & 0.7448 / 0.9003 & 0.7055 / 0.8159 \\
        \bottomrule
    \end{tabular}
    \end{small}
\end{table*}

\subsection{Experimental Results}
Table~\ref{tab:comparison} compares our proposed \texttt{SLUNet} (signed Lanczos Unrolled Network) with the baseline methods in terms of F1 score and accuracy across the four crops. \texttt{SLUNet} achieves the highest F1 scores for soybeans, spring wheat, and grassland/pasture, and the second-highest for corn. 
Its accuracy ranks second for corn and is competitive for the other crops.

The \texttt{Combinatorial Laplacian}, \texttt{Signed Laplacian}, and \texttt{Hybrid Laplacian} \cite{eftekhar2026crop} baselines use the contrastive learning objective for training but do not employ Lanczos unrolling. 
Instead, they rely on the standard eigenvalue solver provided by \texttt{scipy.sparse.linalg.eigsh} for eigenvector computation. 
These baselines serve as ablations of our approach. 
We observe that \texttt{SLUNet} consistently outperforms \texttt{Hybrid Laplacian}, showing the benefit of combining Lanczos unrolling with end-to-end learning.

Deep learning baselines employ substantially larger models and offer less explicit interpretability, reflecting their general-purpose semantic segmentation design and their emphasis on capturing broader spatial context. 
In particular, MA-Net \cite{MAnet} exemplifies this design: its encoder extracts features at multiple resolutions, while its decoder uses attention mechanisms to aggregate information and recover the original resolution. 
However, these architectural components may be unnecessarily complex for our task. 
Extracting features at multiple resolutions increases model size, while aggregating broader spatial context through attention may offer limited benefits for binary crop classification, where the label of a pixel is determined largely by its local neighborhood.

In contrast, \texttt{SLUNet} uses only 945,512 parameters for corn, soybeans, and grassland/pasture, and 945,502 parameters for spring wheat, which requires 15 rather than 20 Lanczos layers. These parameter counts represent approximately $0.6\%$ of the parameters in MA-Net. 
This parameter efficiency is achieved by explicitly modeling pixel similarities and dissimilarities through a learned near-balanced signed graph, and leveraging the efficient unrolled Lanczos algorithm for spectral inference. 
Moreover, the combination of an explicit graph representation and unrolled Lanczos algorithm provides \texttt{SLUNet} with an interpretable architecture.


\section{Conclusion}
\label{sec:conclude}

We proposed a signed graph spectral network for binary crop classification in satellite imagery. 
Our framework learns a near-balanced signed graph from image data and unrolls a Lanczos-based eigensolver to compute the first eigenvector of a hybrid graph Laplacian, whose sign pattern is used for binary label assignment. 
The central observation is that, while this eigenvector exactly recovers the underlying partition for a balanced signed graph, learned feature graphs are typically only near-balanced. 
To address this challenge, we relax the classical Lanczos orthogonality constraints and introduce learnable parameters, enabling the eigensolver to adapt to the classification objective in a data-driven manner. 
Guided by the developed theory, we design regularization terms that promote accurate eigenvector approximation while training the network for end-to-end binary classification.
Experiments on Harmonized Landsat-Sentinel imagery demonstrate classification performance comparable to state-of-the-art deep learning models, while using up to two orders of magnitude fewer parameters.

\section*{Acknowledgment}
The authors used OpenAI GPT-6 for language editing and grammar polishing of the manuscript.

\appendix

\subsection{Proof of Lemma\;\ref{lemma:alpha_i}}
\label{append:proof_lemma1}

\begin{proof} 
Denote by $\u_1$ the eigenvector of $\cL$ corresponding to $\lambda_1$. 
For each Lanczos vector $\v_i$, write 
\begin{align}
\v_i = q_i \u_1 + \z_i,
\end{align}
where $\z_i \perp \u_1$. 
Since $\lambda_2$ is the smallest eigenvalue of $\cL$ restricted to the orthogonal complement of $\u_1$, we have 
\begin{align}
\z_i^\top \cL \z_i \geq \lambda_2 \|\z_i\|_2^2 . 
\end{align}
Using $\|\u_1\|_2 = 1$ and $\|\v_i\|_2=1$,
\begin{align}
\alpha_i = \v_i^\top \cL \v_i = \lambda_1 q_i^2 + \z_i^\top \cL \z_i \geq \lambda_1 q_i^2 + \lambda_2 (1-q_i^2).    
\end{align}
Combining this with the assumption $\alpha_i \leq \lambda_2-\epsilon$, yields
\begin{align}
\lambda_2 - (\lambda_2 - \lambda_1) q_i^2 \leq \lambda_2-\epsilon,    
\end{align}
or equivalently
\begin{align}
q_i^2 \geq \frac{\epsilon}{\lambda_2 - \lambda_1}.    
\end{align}
Since the Lanczos vectors are orthonormal, 
\begin{align}
\sum_{i=1}^m q_i^2 = \sum_{i=1}^m |\u_1^\top \v_i|^2 = \u_1^\top \left( \sum_{i=1}^m \v_i \v_i^\top \right) \u_1.    
\end{align}
Because $\sum_{i=1}^m \v_i \v_i^\top$ is the orthogonal projector onto the Krylov subspace, its eigenvalues are at most $1$.
Therefore,
\begin{align}
\sum_{i=1}^m q_i^2 \leq 1.    
\end{align}
Using the lower bound on each $q_i^2$, 
\begin{align}
m \frac{\epsilon}{\lambda_2-\lambda_1} \leq \sum_{i=1}^m q_i^2 \leq 1.    
\end{align}
Hence, $m \leq \frac{\lambda_2 - \lambda_1}{\epsilon}$.
\end{proof}

\subsection{Proof of Lemma\;\ref{lemma:beta_i}}
\label{append:proof_lemma2}

\begin{proof} 
Since a symmetric tridiagonal matrix is similar, via a diagonal signature matrix, to the matrix obtained by replacing each $\beta_i$ with $|\beta_i|$, the eigenvalues depend only on $|\beta_i|$. Therefore, without loss of generality, we may assume $\beta_i \ge 0$ for all $i$.
Now, assume 
\begin{align}
\beta_i \ge \beta_{\min}, \qquad i=1,\ldots,m-1 . 
\end{align}
Write $\T^o_m = \T_{\min} + \R$, where 
\begin{align}
\R = \sum_{i=1}^{m-1} (\beta_i-\beta_{\min}) \left( e_i e_{i+1}^{\top} + e_{i+1} e_i^{\top} \right).    
\end{align}
The eigenvalues of the uniform path matrix $\T_{\min}$ are
\begin{align}
\mu_k &= 2\beta_{\min} \cos\!\left( \frac{k\pi}{m+1} \right), & k=1,\ldots,m. 
\end{align}
Since the smallest eigenvalue corresponds to $k=m$, denote its unit-norm eigenvector by $\widetilde \y_1$. Its entries are
\begin{align}
\widetilde y_1(j) &= \sqrt{\frac{2}{m+1}}\sin\!\left( \frac{jm\pi}{m+1} \right) \nonumber \\
&= \sqrt{\frac{2}{m+1}}(-1)^{j+1} \sin\!\left( \frac{j\pi}{m+1} \right).    
\end{align}
Therefore, 
\begin{align}
\widetilde y_1(i)\, \widetilde y_1(i+1) < 0, \qquad i=1,\ldots,m-1.    
\end{align}
Using the Rayleigh characterization of the smallest eigenvalue, 
\begin{align}
\theta_1(\T^o_m) = \min_{\|x\|_2=1} \x^{\top} \T^o_m \x \le \widetilde \y_1^{\top} \T^o_m \widetilde \y_1 .    
\end{align}
Further, 
\begin{align}
\widetilde \y_1^{\top} \T^o_m \widetilde \y_1 = \theta_1(\T_{\min}) + 2 \sum_{i=1}^{m-1} (\beta_i-\beta_{\min}) \widetilde y_1(i)\widetilde y_1(i+1).    
\end{align}
Since $\beta_i-\beta_{\min}\ge 0$ and $ \widetilde y_1(i)\widetilde y_1(i+1)<0$, every term in the summation is non-positive. 
Hence, 
\begin{align}
\widetilde y_1^{\top} \T^o_m \widetilde y_1 \le \theta_1(\T_{\min}).    
\end{align}
Combining the above inequalities yields
\begin{align}
\theta_1(\T^o_m) \le \theta_1(\T_{\min}),    
\end{align}
which completes the proof. 
\end{proof}

\subsection{Proof of Theorem\;\ref{thm:alpha_beta}}
\label{append:proof_thm}

\begin{proof} 
(i) By Lemma\;\ref{lemma:alpha_i}, 
\begin{align}
|\u_1^\top \v_i|^2 \ge \frac{\epsilon}{\lambda_2 - \lambda_1}, \qquad i=1,\ldots,m.    
\end{align}
Let $\P_m = \V_m \V_m^\top$ denote the orthogonal projector onto $\cK_m$. 
Then 
\begin{align}
\u_1^\top \P_m \u_1 = \sum_{i=1}^{m} |\u_1^\top \v_i|^2 \ge \frac{m\epsilon}{\lambda_2 - \lambda_1}.
\end{align}
Using 
\begin{align}
\sin^2\angle(\u_1,\cK_m) = 1-\u_1^\top \P_m \u_1,
\end{align}
we obtain 
\begin{align}
\sin^2\angle(\u_1,\cK_m) \le 1-\frac{m\epsilon}{\lambda_2 - \lambda_1}.
\end{align}
Taking square roots yields 
\begin{align}
\sin\angle(\u_1,\cK_m) \le \sqrt{ 1-\frac{m\epsilon}{\lambda_2 - \lambda_1} }.    
\end{align}

(ii) Define 
\begin{align}
\T_o = \begin{bmatrix} 0 & \beta_1 \\ \beta_1 & 0 & \beta_2\\ & \beta_2 & \ddots & \ddots\\ && \ddots & 0 & \beta_{m-1}\\ &&& \beta_{m-1} & 0 \end{bmatrix}.
\end{align}
Then $\T_m = \operatorname{diag}(\alpha_1,\ldots,\alpha_m) + \T_o$.  
Since \[ \alpha_i\le \lambda_2-\epsilon, \] we have 
\begin{align}
\operatorname{diag}(\alpha_1,\ldots,\alpha_m) \preceq (\lambda_2-\epsilon) \I.    
\end{align}
Therefore, 
\begin{align}
\T_m \preceq (\lambda_2-\epsilon) \I + \T_o.    
\end{align}
By eigenvalue monotonicity, 
\begin{align}
\theta_1(\T_m) \le \theta_1\!\left( (\lambda_2-\epsilon)\I + \T_o \right).    
\end{align}
Since adding a scaled identity matrix shifts all eigenvalues by the same amount, 
\begin{align}
\theta_1\!\left( (\lambda_2-\epsilon) \I + \T_o \right) = (\lambda_2-\epsilon)+\theta_1(\T_o).    
\end{align}
By Lemma\;\ref{lemma:beta_i}, 
\begin{align}
\theta_1(\T_o) \le -2\beta_{\min} \cos\!\left( \frac{\pi}{m+1} \right).
\end{align}
Hence 
\begin{align}
\theta_1(\T_m) \le (\lambda_2-\epsilon) - 2\beta_{\min} \cos\!\left( \frac{\pi}{m+1} \right).
\end{align}
Finally, because $\T_m = \V_m^\top \cL \V_m$ is a Rayleigh-Ritz projection of the symmetric matrix $\cL$, the eigenvalue interlacing theorem implies 
\begin{align}
\lambda_1(\cL)\le \theta_1(\T_m). 
\end{align}

\noindent (iii) Since $\beta_{\min}>0$, every off-diagonal entry of $\T_m$ is nonzero. 
Hence, $\T_m$ is an unreduced symmetric tridiagonal matrix, and its eigenvalues are simple.
To see this, let $\T_m \y = \theta \y$ with $\y \neq \0$. 
The first row gives $y_2 = (\theta-\alpha_1)y_1/\beta_1$, and row $j$ gives 
\begin{align}
y_{j+1} = \frac{(\theta-\alpha_j)y_j - \beta_{j-1}y_{j-1}}{\beta_j}, \qquad j=2,\ldots,m-1.
\end{align}
Thus, $y_2,\ldots,y_m$ are determined by $y_1$, so the eigenspace for $\theta$ is one-dimensional. 
Since $\T_m$ is symmetric, $\theta$ is a simple eigenvalue. 
In particular, $\theta_1(\T_m)$ is simple, and its unit-norm eigenvector $\y_1$ is unique up to sign. 
The Ritz vector $\x_1 = \V_m \y_1$ is therefore well defined up to sign, and $|\u_1^\top \x_1|$ does not depend on the choice of sign.

Because $\V_m$ has orthonormal columns and $\|\y_1\|_2 = 1$, 
\begin{align}
\|\x_1\|_2^2 = \y_1^\top \V_m^\top \V_m \y_1 = \|\y_1\|_2^2 = 1,
\end{align}
and, using $\T_m = \V_m^\top \cL \V_m$,
\begin{align}
\x_1^\top \cL \x_1 = \y_1^\top \V_m^\top \cL \V_m \y_1 = \y_1^\top \T_m \y_1 = \theta_1(\T_m). 
\end{align}
Write 
\begin{align}
\x_1 = q\,\u_1 + \z, \qquad q = \u_1^\top \x_1, \quad \z \perp \u_1.
\end{align}
Then, $\|\z\|_2^2 = 1 - q^2$. 
Since $\u_1$ is an eigenvector of $\cL$, the cross terms vanish, and since $\lambda_2$ is the smallest eigenvalue of $\cL$ restricted to the orthogonal complement of $\u_1$, we have $\z^\top \cL \z \ge \lambda_2 \|\z\|_2^2$. 
Therefore,
\begin{align}
\theta_1(\T_m) = \x_1^\top \cL \x_1 = \lambda_1 q^2 + \z^\top \cL \z &\ge \lambda_1 q^2 + \lambda_2 (1-q^2) 
\nonumber \\ 
&= \lambda_2 - (\lambda_2-\lambda_1)q^2.
\end{align}
Rearranging yields
\begin{align}
|\u_1^\top \x_1|^2 = q^2 \ge \frac{\lambda_2 - \theta_1(\T_m)}{\lambda_2-\lambda_1}.
\end{align}
By part (ii), 
\begin{align}
\lambda_2 - \theta_1(\T_m) \ge \epsilon + 2\beta_{\min}\cos\!\left(\frac{\pi}{m+1}\right) > 0,
\end{align}
and so
\begin{align}
|\u_1^\top \x_1|^2 \ge \frac{\epsilon + 2\beta_{\min}\cos\!\left(\frac{\pi}{m+1}\right)}{\lambda_2-\lambda_1}.
\end{align}
Finally, since $\|\x_1\|_2 = \|\u_1\|_2 = 1$, we have $\sin^2\angle(\u_1,\x_1) = 1 - |\u_1^\top \x_1|^2$, which gives
\begin{align}
\sin^2\angle(\u_1,\x_1) \le 1 - \frac{\epsilon + 2\beta_{\min}\cos\!\left(\frac{\pi}{m+1}\right)}{\lambda_2-\lambda_1}.
\end{align}
This completes the proof.
\end{proof}

\subsection{Proof of Lemma\;\ref{lemma:real_evalue}}
\label{append:proof_lemma3}

\begin{proof} 
Define the diagonal matrix $\Delta = \mathrm{diag}(\kappa_1,\ldots,\kappa_m)$ by
$\kappa_1=1$ and $\kappa_j=\kappa_{j-1}\ell_j/\sqrt{\ell_j u_{j-1}}$ for $j=2,\ldots,m$.
Since $\ell_j u_{j-1}>0$, every $\kappa_j$ is real and nonzero. 
Let $\S=\Delta^{-1}\T_m\Delta$.
The diagonal entries of $\S$ equal those of $\T_m$, while for $j=2,\ldots,m$, 
$S_{j,j-1}=\ell_j\kappa_{j-1}/\kappa_j=\sqrt{\ell_j u_{j-1}}$
and $S_{j-1,j}=u_{j-1}\kappa_j/\kappa_{j-1}=\sqrt{\ell_j u_{j-1}}$. Hence, $\S$ is symmetric:
\begin{small}
\begin{align}
\S = \left[ \begin{array}{ccccc} a_1 & \sqrt{\ell_2u_1} & & & 0 \\ \sqrt{\ell_2u_1} & a_2 & \sqrt{\ell_3u_2} & & \\ & \sqrt{\ell_3u_2} & \ddots & \ddots & \\ & & \ddots & a_{m-1} & \sqrt{\ell_mu_{m-1}} \\ 0 & & & \sqrt{\ell_mu_{m-1}} & a_m \end{array} \right].
\end{align}
\end{small}\noindent
Since $\T_m$ is similar to the real symmetric matrix $\S$, $\T_m$ and $\S$ have the same eigenvalues. Because every real symmetric matrix has real eigenvalues, all eigenvalues of $\T_m$ are real. 
\end{proof}


\bibliographystyle{ieeetr}   
\bibliography{ref}

@book{richards2022remote,
  title={Remote sensing digital image analysis},
  author={Richards, John A and Richards, John A and others},
  volume={5},
  year={2022},
  publisher={Springer}
}

@article{boryan2011monitoring,
  title={Monitoring US agriculture: the US department of agriculture, national agricultural statistics service, cropland data layer program},
  author={Boryan, Claire and Yang, Zhengwei and Mueller, Rick and Craig, Mike},
  journal={Geocarto International},
  volume={26},
  number={5},
  pages={341--358},
  year={2011},
  publisher={Taylor \& Francis}
}

@inproceedings{eftekhar2026crop,
  title={CROP Classification in Satellite Images via First Eigenvector of Learned Signed Graph Laplacian},
  author={Eftekhar, Parham and Cheung, Gene and Eadie, Tim},
  booktitle={ICASSP 2026-2026 IEEE International Conference on Acoustics, Speech and Signal Processing (ICASSP)},
  pages={601--605},
  year={2026},
  organization={IEEE}
}

@article{von2007tutorial,
  title={A tutorial on spectral clustering},
  author={Von Luxburg, Ulrike},
  journal={Statistics and computing},
  volume={17},
  number={4},
  pages={395--416},
  year={2007},
  publisher={Springer}
}

@inproceedings{radford2021learning,
  title={Learning transferable visual models from natural language supervision},
  author={Radford, Alec and Kim, Jong Wook and Hallacy, Chris and Ramesh, Aditya and Goh, Gabriel and Agarwal, Sandhini and Sastry, Girish and Askell, Amanda and Mishkin, Pamela and Clark, Jack and others},
  booktitle={International conference on machine learning},
  pages={8748--8763},
  year={2021},
  organization={PMLR}
}

@inproceedings{schroff2015facenet,
  title={Facenet: A unified embedding for face recognition and clustering},
  author={Schroff, Florian and Kalenichenko, Dmitry and Philbin, James},
  booktitle={Proceedings of the IEEE conference on computer vision and pattern recognition},
  pages={815--823},
  year={2015}
}

@inproceedings{hadsell2006dimensionality,
  title={Dimensionality reduction by learning an invariant mapping},
  author={Hadsell, Raia and Chopra, Sumit and LeCun, Yann},
  booktitle={2006 IEEE computer society conference on computer vision and pattern recognition (CVPR'06)},
  volume={2},
  pages={1735--1742},
  year={2006},
  organization={IEEE}
}

@article{allwein2000reducing,
  title={Reducing multiclass to binary: A unifying approach for margin classifiers},
  author={Allwein, Erin L and Schapire, Robert E and Singer, Yoram},
  journal={Journal of machine learning research},
  volume={1},
  number={Dec},
  pages={113--141},
  year={2000}
}

@misc{goodfellow2016deep,
  title={Deep learning},
  author={Goodfellow, Ian},
  year={2016},
  publisher={MIT press}
}

@article{xie2021segformer,
  title={SegFormer: Simple and efficient design for semantic segmentation with transformers},
  author={Xie, Enze and Wang, Wenhai and Yu, Zhiding and Anandkumar, Anima and Alvarez, Jose M and Luo, Ping},
  journal={Advances in neural information processing systems},
  volume={34},
  pages={12077--12090},
  year={2021}
}

@article{chen2017deeplab,
  title={Deeplab: Semantic image segmentation with deep convolutional nets, atrous convolution, and fully connected crfs},
  author={Chen, Liang-Chieh and Papandreou, George and Kokkinos, Iasonas and Murphy, Kevin and Yuille, Alan L},
  journal={IEEE transactions on pattern analysis and machine intelligence},
  volume={40},
  number={4},
  pages={834--848},
  year={2017},
  publisher={IEEE}
}

@article{knyazev2001toward,
  title={Toward the optimal preconditioned eigensolver: Locally optimal block preconditioned conjugate gradient method},
  author={Knyazev, Andrew V},
  journal={SIAM journal on scientific computing},
  volume={23},
  number={2},
  pages={517--541},
  year={2001},
  publisher={SIAM}
}

@article{green2018usa,
  title={Where is the USA Corn Belt, and how is it changing?},
  author={Green, Timothy R and Kipka, Holm and David, Olaf and McMaster, Gregory S},
  journal={Science of the Total Environment},
  volume={618},
  pages={1613--1618},
  year={2018},
  publisher={Elsevier}
}

@article{lecun1998gradient,
  title={Gradient-based learning applied to document recognition},
  author={LeCun, Yann and Bottou, L{\'e}on and Bengio, Yoshua and Haffner, Patrick},
  journal={Proceedings of the IEEE},
  volume={86},
  number={11},
  pages={2278--2324},
  year={1998},
  publisher={Ieee}
}

@article{monitoring,
  title={ARSET - Satellite Remote Sensing for Agricultural Applications. NASA Applied Remote Sensing Training Program (ARSET). http://appliedsciences.nasa.gov/get-involved/training/english/arset-satellite-remote-sensing-agricultural-applications}
}

@article{craig2010history,
  title={A history of the cropland data layer at NASS},
  author={Craig, Michael},
  journal={USDA NASS CropScape},
  year={2010}
}

@article{claverie2018harmonized,
  title={The Harmonized Landsat and Sentinel-2 surface reflectance data set},
  author={Claverie, Martin and Ju, Junchang and Masek, Jeffrey G and Dungan, Jennifer L and Vermote, Eric F and Roger, Jean-Claude and Skakun, Sergii V and Justice, Christopher},
  journal={Remote sensing of environment},
  volume={219},
  pages={145--161},
  year={2018},
  publisher={Elsevier}
}

@ARTICLE{shi00,
  author={Jianbo Shi and Malik, J.},
  journal={IEEE Transactions on Pattern Analysis and Machine Intelligence}, 
  title={Normalized cuts and image segmentation}, 
  year={2000},
  volume={22},
  number={8},
  pages={888-905},
  doi={10.1109/34.868688}}

@article{monga2021algorithm,
  title={Algorithm unrolling: Interpretable, efficient deep learning for signal and image processing},
  author={Monga, Vishal and Li, Yuelong and Eldar, Yonina C},
  journal={IEEE Signal Processing Magazine},
  volume={38},
  number={2},
  pages={18--44},
  year={2021},
  publisher={IEEE}
}

@article{lanczos1950iteration,
  title={An iteration method for the solution of the eigenvalue problem of linear differential and integral operators},
  author={Lanczos, Cornelius},
  journal={Journal of research of the National Bureau of Standards},
  volume={45},
  number={4},
  pages={255--282},
  year={1950}
}

@inproceedings{rustowicz2019semantic,
  title={Semantic Segmentation of Crop Type in Africa: A Novel Dataset and Analysis of Deep Learning Methods.},
  author={Rustowicz, Rose M and Cheong, Robin and Wang, Lijing and Ermon, Stefano and Burke, Marshall and Lobell, David B},
  booktitle={CVPR Workshops},
  volume={2},
  pages={8},
  year={2019}
}

@article{thomas1949elliptic,
  title={Elliptic problems in linear difference equations over a network},
  author={Thomas, Llewellyn Hilleth},
  journal={Watson Sci. Comput. Lab. Rept., Columbia University, New York},
  volume={1},
  number={71},
  pages={71--72},
  year={1949}
}

@BOOK(varga04,
  TITLE = "{Gershgorin} and his circles",
  AUTHOR = "Varga, R. S.",
  PUBLISHER = "Springer",
  YEAR = "2004")

@article{harary1953notion,
  title={On the notion of balance of a signed graph.},
  author={Harary, Frank},
  journal={Michigan Mathematical Journal},
  volume={2},
  number={2},
  pages={143--146},
  year={1953},
  publisher={University of Michigan, Department of Mathematics}
}

@article{hou2003laplacian,
  title={On the Laplacian eigenvalues of signed graphs},
  author={Hou, Yaoping and Li, Jiongsheng and Pan, Yongliang},
  journal={Linear and Multilinear Algebra},
  volume={51},
  number={1},
  pages={21--30},
  year={2003},
  publisher={Taylor \& Francis}
}

@article{knyazev2017spectral,
  title={On spectral partitioning of signed graphs},
  author={Knyazev, Andrew V},
  journal={arXiv preprint arXiv:1701.01394},
  year={2017}
}

@inproceedings{mercado2019spectral,
  title={Spectral clustering of signed graphs via matrix power means},
  author={Mercado, Pedro and Tudisco, Francesco and Hein, Matthias},
  booktitle={International Conference on Machine Learning},
  pages={4526--4536},
  year={2019},
  organization={PMLR}
}

@ARTICLE{pang17,
  author={Pang, Jiahao and Cheung, Gene},
  journal={IEEE Transactions on Image Processing},
  title={Graph {Laplacian} Regularization for Image Denoising: Analysis in the Continuous Domain},
  year={2017},
  volume={26},
  number={4},
  pages={1770-1785},
  doi={10.1109/TIP.2017.2651400}}

@INPROCEEDINGS(cheung18,
  TITLE = "Graph Spectral Image Processing",
  AUTHOR = "G. Cheung and E. Magli and Y. Tanaka and M. Ng",
  BOOKTITLE = "Proceedings of the {IEEE}",
  VOLUME = "106, no.5",
  PAGES = "907-930",
  MONTH = "May",
  YEAR = "2018")

@INPROCEEDINGS(ortega18ieee,
  TITLE = "Graph Signal Processing: Overview, Challenges, and Applications",
  AUTHOR = "A. Ortega and P. Frossard and J. Kovacevic and J. M. F. Moura and P. Vandergheynst",
  BOOKTITLE = "Proceedings of the {IEEE}",
  VOLUME = "106, no.5",
  PAGES = "808-828",
  MONTH = "May",
  YEAR = "2018")

@article{bai19,
  TITLE = "Graph-Based Blind Image Deblurring from a Single Photograph",
  AUTHOR = "Y. Bai and G. Cheung and X. Liu and W. Gao",
   journal = "IEEE Transactions on Image Processing",
  VOLUME = "28, no.3",
  PAGES = "1404-1418",
  YEAR = "2019"}

@ARTICLE{monga21,
  author={Monga, Vishal and Li, Yuelong and Eldar, Yonina C.},
  journal={IEEE Signal Processing Magazine},
  title={Algorithm Unrolling: Interpretable, Efficient Deep Learning for Signal and Image Processing},
  year={2021},
  volume={38},
  number={2},
  pages={18-44},
  doi={10.1109/MSP.2020.3016905}}

@ARTICLE{yang22,
  author={Yang, Cheng and Cheung, Gene and Hu, Wei},
  journal={IEEE Transactions on Pattern Analysis and Machine Intelligence},
  title={Signed Graph Metric Learning via {Gershgorin} Disc Perfect Alignment},
  year={2022},
  volume={44},
  number={10},
  pages={7219-7234},
  doi={10.1109/TPAMI.2021.3091682}}

@ARTICLE{chen24,
  author={Chen, Fei and Cheung, Gene and Zhang, Xue},
  journal={IEEE Transactions on Signal Processing}, 
  title={Manifold Graph Signal Restoration Using Gradient Graph {Laplacian} Regularizer},
  year={2024},
  volume={72},
  number={},
  pages={744-761},
  doi={10.1109/TSP.2023.3343560}}

@ARTICLE{dinesh24,
  author={Dinesh, Chinthaka and Cheung, Gene and Bagheri, Saghar and Bajic, Ivan V.},
  journal={IEEE Transactions on Pattern Analysis and Machine Intelligence},
  title={Efficient Signed Graph Sampling via Balancing \& {Gershgorin} Disc Perfect Alignment},
  year={2024},
  volume={},
  number={},
  pages={1-18},
  doi={10.1109/TPAMI.2024.3524180}}

@inproceedings{Kunegis2010,
  author    = {J{\'e}r{\^o}me Kunegis and Stephan Schmidt and Andreas Lommatzsch and
               J{\"u}rgen Lerner and Ernesto W. De Luca and Sahin Albayrak},
  title     = {Spectral Analysis of Signed Graphs for Clustering, Prediction and Visualization},
  booktitle = {Proceedings of the 2010 SIAM International Conference on Data Mining (SDM)},
  pages      = {559--570},
  year       = {2010},
  doi        = {10.1137/1.9781611972801.49},
  url        = {https://epubs.siam.org/doi/abs/10.1137/1.9781611972801.49}
}

@book{saad2011numerical,
  title={Numerical methods for large eigenvalue problems: revised edition},
  author={Saad, Yousef},
  year={2011},
  publisher={SIAM}
}

@article{leon2013gram,
  title={Gram-Schmidt orthogonalization: 100 years and more},
  author={Leon, Steven J and Bj{\"o}rck, {\\AA}ke and Gander, Walter},
  journal={Numerical Linear Algebra with Applications},
  volume={20},
  number={3},
  pages={492--532},
  year={2013},
  publisher={Wiley Online Library}
}

@article{bresson2013multiclass,
  title={Multiclass total variation clustering},
  author={Bresson, Xavier and Laurent, Thomas and Uminsky, David and Von Brecht, James},
  journal={Advances in neural information processing systems},
  volume={26},
  year={2013}
}

@article{thuc2024interpretable,
  title={Interpretable lightweight transformer via unrolling of learned graph smoothness priors},
  author={Thuc, Tam and Eftekhar, Parham and Hosseini, Seyed A and Cheung, Gene and Chou, Philip A},
  journal={Advances in Neural Information Processing Systems},
  volume={37},
  pages={6393--6416},
  year={2024}
}

@article{qi2025lightweight,
  title={Lightweight and Interpretable Transformer via Mixed Graph Algorithm Unrolling for Traffic Forecast},
  author={Qi, Ji and Do, Tam Thuc and Liu, Mingxiao and Pan, Zhuoshi and Li, Yuzhe and Cheung, Gene and Zhao, H Vicky},
  journal={arXiv preprint arXiv:2505.13102},
  year={2025}
}

@article{cai2025unrolling,
  title={Unrolling plug-and-play gradient graph Laplacian regularizer for image restoration},
  author={Cai, Jianghe and Cheung, Gene and Chen, Fei},
  journal={IEEE Transactions on Image Processing},
  volume={35},
  pages={1817--1831},
  year={2025},
  publisher={IEEE}
}

@inproceedings{yao2026lightweight,
  title={Lightweight Transformer for EEG Classification via Balanced Signed Graph Algorithm Unrolling},
  author={Yao, Junyi and Eftekhar, Parham and Cheung, Gene and Liu, Xujin and Wang, Yao and Hu, Wei},
  booktitle={International Conference on Learning Representations},
  volume={2026},
  pages={149466--149486},
  year={2026}
}

@article{yu2023white,
  title={White-box transformers via sparse rate reduction},
  author={Yu, Yaodong and Buchanan, Sam and Pai, Druv and Chu, Tianzhe and Wu, Ziyang and Tong, Shengbang and Haeffele, Benjamin and Ma, Yi},
  journal={Advances in Neural Information Processing Systems},
  volume={36},
  pages={9422--9457},
  year={2023}
}

@article{MAnet,
  title={Ma-net: A multi-scale attention network for liver and tumor segmentation},
  author={Fan, Tongle and Wang, Guanglei and Li, Yan and Wang, Hongrui},
  journal={Ieee Access},
  volume={8},
  pages={179656--179665},
  year={2020},
  publisher={IEEE}
}

@article{chen2019rethinking,
  title={Rethinking atrous convolution for semantic image segmentation. arXiv 2017},
  author={Chen, Liang-Chieh and Papandreou, George and Schroff, Florian and Adam, Hartwig},
  journal={arXiv preprint arXiv:1706.05587},
  volume={2},
  pages={1},
  year={2019}
}

@inproceedings{long2015fully,
  title={Fully convolutional networks for semantic segmentation},
  author={Long, Jonathan and Shelhamer, Evan and Darrell, Trevor},
  booktitle={Proceedings of the IEEE conference on computer vision and pattern recognition},
  pages={3431--3440},
  year={2015}
}

@article{li2018pyramid,
  title={Pyramid attention network for semantic segmentation},
  author={Li, Hanchao and Xiong, Pengfei and An, Jie and Wang, Lingxue},
  journal={arXiv preprint arXiv:1805.10180},
  year={2018}
}

@inproceedings{zhao2017pyramid,
  title={Pyramid scene parsing network},
  author={Zhao, Hengshuang and Shi, Jianping and Qi, Xiaojuan and Wang, Xiaogang and Jia, Jiaya},
  booktitle={Proceedings of the IEEE conference on computer vision and pattern recognition},
  pages={2881--2890},
  year={2017}
}

@article{kirillov2017unified,
  title={A unified architecture for instance and semantic segmentation (2017)},
  author={Kirillov, A and He, K and Girshick, R and Doll{\'a}r, P},
  journal={SF Banu et al./WEU-Net: A Weight Excitation U-Net for Lung Nodule Segmentation},
  volume={356},
  year={2017}
}

@inproceedings{UPerNet,
  title={Unified perceptual parsing for scene understanding},
  author={Xiao, Tete and Liu, Yingcheng and Zhou, Bolei and Jiang, Yuning and Sun, Jian},
  booktitle={Proceedings of the European conference on computer vision (ECCV)},
  pages={418--434},
  year={2018}
}

@inproceedings{matz2020learning,
  title={Learning signed graphs from data},
  author={Matz, Gerald and Dittrich, Thomas},
  booktitle={ICASSP 2020-2020 IEEE International Conference on Acoustics, Speech and Signal Processing (ICASSP)},
  pages={5570--5574},
  year={2020},
  organization={IEEE}
}

@article{karaaslanli2022scsgl,
  title={scSGL: kernelized signed graph learning for single-cell gene regulatory network inference},
  author={Karaaslanli, Abdullah and Saha, Satabdi and Aviyente, Selin and Maiti, Tapabrata},
  journal={Bioinformatics},
  volume={38},
  number={11},
  pages={3011--3019},
  year={2022},
  publisher={Oxford University Press}
}
\end{document}


\section{Additional findings}

\subsection{Nullspace Components of the Lanczos Vectors}

Let $\cL \in \mathbb{R}^{N\times N}$ be a symmetric positive
semidefinite matrix with a one-dimensional nullspace. Let $\u_1$
denote the normalized eigenvector corresponding to the zero
eigenvalue, i.e.,
\begin{equation}
\cL \u_1 = \0,
\qquad
\|\u_1\|_2 = 1.
\end{equation}
Let $\{\v_i\}_{i=1}^m$ denote the orthonormal Lanczos vectors
generated from a unit-norm initial vector $\v_1$. Define
\begin{equation}
c_i \triangleq \u_1^\top \v_i.
\end{equation}
Thus, $c_i$ represents the component of the $i$-th Lanczos vector
along the nullspace direction $\u_1$.

The Lanczos three-term recurrence is
\begin{equation}
\cL \v_i
=
\beta_{i-1}\v_{i-1}
+
\alpha_i\v_i
+
\beta_i\v_{i+1},
\end{equation}
where $\beta_0=0$. Multiplying both sides by $\u_1^\top$ and using
$\u_1^\top\cL=\0^\top$ gives
\begin{equation}
\beta_{i-1}c_{i-1}
+
\alpha_i c_i
+
\beta_i c_{i+1}
=
0.
\end{equation}
Therefore,
\begin{equation}
\boxed{
c_{i+1}
=
-\frac{\alpha_i c_i+\beta_{i-1}c_{i-1}}{\beta_i}.
}
\label{eq:c_recurrence}
\end{equation}

For $i=1$, since $\beta_0=0$, we obtain
\begin{equation}
\boxed{
c_2
=
-\frac{\alpha_1}{\beta_1}c_1.
}
\label{eq:c2}
\end{equation}

Applying \eqref{eq:c_recurrence} recursively gives
\begin{align}
c_3
&=
-\frac{\alpha_2c_2+\beta_1c_1}{\beta_2}
\nonumber\\
&=
\frac{\alpha_1\alpha_2-\beta_1^2}
{\beta_1\beta_2}c_1,
\end{align}
and
\begin{align}
c_4
&=
-\frac{\alpha_3c_3+\beta_2c_2}{\beta_3}
\nonumber\\
&=
-\frac{
\alpha_1\alpha_2\alpha_3
-\alpha_1\beta_2^2
-\alpha_3\beta_1^2
}{
\beta_1\beta_2\beta_3
}c_1.
\end{align}

These expressions admit a compact representation in terms of the
leading principal submatrix of the Lanczos tridiagonal matrix,
\begin{equation}
\T_i =
\begin{bmatrix}
\alpha_1 & \beta_1 & & & \\
\beta_1 & \alpha_2 & \ddots & & \\
& \ddots & \ddots & \ddots & \\
& & \ddots & \alpha_{i-1} & \beta_{i-1} \\
& & & \beta_{i-1} & \alpha_i
\end{bmatrix}.
\end{equation}

In particular, the coefficients satisfy
\begin{equation}
\boxed{
c_{i+1}
=
(-1)^i
\frac{\det(\T_i)}
{\displaystyle\prod_{j=1}^{i}\beta_j}
c_1.
}
\label{eq:c_det}
\end{equation}

Let $\theta_1^{(i)},\ldots,\theta_i^{(i)}$ denote the eigenvalues
(Ritz values) of $\T_i$. Since
\begin{equation}
\det(\T_i)
=
\prod_{j=1}^{i}\theta_j^{(i)},
\end{equation}
\eqref{eq:c_det} can equivalently be written as
\begin{equation}
\boxed{
c_{i+1}
=
(-1)^i
\frac{
\displaystyle\prod_{j=1}^{i}\theta_j^{(i)}
}{
\displaystyle\prod_{j=1}^{i}\beta_j
}
c_1.
}
\label{eq:c_ritz}
\end{equation}

This relation shows that the nullspace component $c_{i+1}$ is
directly related to the Ritz values of $\T_i$. In particular, if
$\T_i$ develops a Ritz value close to zero, then
\begin{equation}
\theta_j^{(i)} \approx 0
\quad\Longrightarrow\quad
\det(\T_i)\approx 0
\quad\Longrightarrow\quad
c_{i+1}\approx 0.
\end{equation}

\subsection{Relation to the Ritz Approximation of the Nullspace}

Let
\begin{equation}
\V_m =
\begin{bmatrix}
\v_1 & \v_2 & \cdots & \v_m
\end{bmatrix},
\end{equation}
and define
\begin{equation}
\c =
\V_m^\top \u_1
=
\begin{bmatrix}
c_1 & c_2 & \cdots & c_m
\end{bmatrix}^{\top}.
\end{equation}

The Lanczos decomposition is
\begin{equation}
\cL\V_m
=
\V_m\T_m
+
\beta_m\v_{m+1}\e_m^\top,
\end{equation}
where $\e_m$ is the $m$-th canonical basis vector.

Since $\cL\u_1=\0$, we obtain
\begin{equation}
\boxed{
\T_m\c
=
-\beta_m c_{m+1}\e_m.
}
\label{eq:Tc}
\end{equation}

Consequently,
\begin{equation}
\boxed{
\|\T_m\c\|_2
=
\beta_m |c_{m+1}|.
}
\label{eq:Tc_norm}
\end{equation}

Equation~\eqref{eq:Tc_norm} provides a direct interpretation of
$c_{m+1}$. Specifically, $c_{m+1}$ determines the residual obtained
when representing the zero-eigenvalue eigenvector $\u_1$ in the
Lanczos subspace.

If
\begin{equation}
c_{m+1}=0,
\end{equation}
then
\begin{equation}
\T_m\c=\0,
\end{equation}
and hence $\c$ is an exact eigenvector of $\T_m$ associated with
the zero eigenvalue. More generally, a small value of
$|c_{m+1}|$ implies a small residual for the corresponding
zero-eigenvalue Ritz approximation.

\subsection{Projection of the Nullspace onto the Krylov Subspace}

The individual coefficients $c_i$ describe how the nullspace
direction $\u_1$ is distributed among the Lanczos basis vectors.
A useful aggregate measure is
\begin{equation}
\eta_m
\triangleq
\sum_{i=1}^{m}c_i^2
=
\|\V_m^\top\u_1\|_2^2.
\end{equation}

Since $\V_m\V_m^\top$ is the orthogonal projector onto the Krylov
subspace
\begin{equation}
\cK_m
=
\operatorname{span}
\{\v_1,\ldots,\v_m\},
\end{equation}
we have
\begin{equation}
\boxed{
\eta_m
=
\|\V_m\V_m^\top\u_1\|_2^2.
}
\label{eq:eta}
\end{equation}

Equivalently,
\begin{equation}
\boxed{
1-\eta_m
=
\left\|
\u_1-\V_m\V_m^\top\u_1
\right\|_2^2.
}
\label{eq:nullspace_distance}
\end{equation}

Thus, $1-\eta_m$ is the squared distance between the nullspace
eigenvector and the $m$-dimensional Krylov subspace.

For a balanced signed graph, $\u_1$ can be replaced by the normalized
signed indicator vector
\begin{equation}
\u_s
=
\frac{1}{\sqrt{N}}
\begin{bmatrix}
\1_{\mathcal V_+}\\
-\1_{\mathcal V_-}
\end{bmatrix},
\end{equation}
where $\mathcal V_+$ and $\mathcal V_-$ denote the two balanced
partitions. In this case,
\begin{equation}
\cL\u_s=\0,
\end{equation}
and
\begin{equation}
c_i
=
\u_s^\top\v_i
\end{equation}
measures the contribution of the balanced signed partition mode
to the $i$-th Lanczos vector.